\documentclass[preprint,prd,tightenlines,nofootinbib,eqsecnum]{revtex4-1}

\usepackage{amsmath}
\usepackage{amsthm}
\usepackage{amsfonts}
\usepackage{amssymb}
\usepackage{mathrsfs}
\usepackage{bm}
\usepackage{graphicx}
\usepackage{hyperref}
\usepackage{color}
\usepackage{array}
\usepackage{booktabs}
\usepackage{verbatim}
\usepackage{addfont}
\usepackage{enumerate}
\usepackage[modulo]{lineno}
\usepackage[scr=boondox,scrscaled=1.05]{mathalfa}
\usepackage{comment}
\usepackage{ulem}

\newcolumntype{C}{>{$}c<{$}}
\AtBeginDocument{
\heavyrulewidth=.08em
\lightrulewidth=.05em
\cmidrulewidth=.03em
\belowrulesep=.65ex
\belowbottomsep=0pt
\aboverulesep=.4ex
\abovetopsep=0pt
\cmidrulesep=\doublerulesep
\cmidrulekern=.5em
\defaultaddspace=.5em
}

\definecolor{CiteColor}{rgb}{0,0.6,0.1}
\definecolor{URLColor}{rgb}{1,0,0.7}
\definecolor{LinksColor}{rgb}{1,0,0.1}

\hypersetup{colorlinks,linkcolor=LinksColor,citecolor=CiteColor,urlcolor=URLColor}

\newcommand{\beq}{\begin{equation}}
\newcommand{\eeq}{\end{equation}}
\newcommand{\ud}{\mathrm{d}}
\newcommand{\ui}{\mathrm{i}}

\newcommand{\RR}{\mathbb{R}}
\newcommand{\II}{\mathbb{I}}
\newcommand{\ZZ}{\mathbb{Z}}
\newcommand{\QQ}{\mathbb{Q}}
\newcommand{\CC}{\mathbb{C}}
\newcommand{\lfk}{\mathfrak{l}}
\newcommand{\bfk}{\mathfrak{b}}
\newcommand{\Bfk}{\mathfrak{B}}

\newcommand{\psiHe}{\psi_{\rm{He}}}
\newcommand{\psiHo}{\psi_{\rm{Ho}}}
\newcommand{\psiHa}{\psi_{\rm{Ha}}}
\newcommand{\psiKe}{\psi_{\rm{Ke}}}
\newcommand{\psiBo}{\psi_{\rm{Bo}}}

\newtheorem*{theorem*}{Theorem}

\begin{document}

\setcounter{tocdepth}{1}

\title{New Methods of Isochrone Mechanics \vspace{1cm}}

\author{Paul Ramond$^{\dag \ddag}$\footnote{Email: paul.ramond@obspm.fr (contact author)} and J\'er$\hat{\text{o}}$me Perez$^{\ddag}$\footnote{Email: jerome.perez@ensta-paris.fr}\vspace{0.5cm}}
\affiliation{$^{\dag}$Laboratoire Univers et TH\'eories\\Observatoire de Paris, PSL Research University, CNRS, Paris University, Sorbonne Paris Cit\'e\\
92190 Meudon, France}
\author{}
\affiliation{$^{\ddag}$Laboratoire de Math\'ematiques Appliqu\'ees\\
 UMA, ENSTA Paris, Institut Polytechnique de Paris,\\91120
Palaiseau, France\vspace{2cm}}

\begin{abstract}

\vspace{5mm}

Isochrone potentials, as defined by Michel Hénon in the fifties, are spherically symmetric potentials within which a particle orbits with a radial period that is independent of its angular momentum. Isochrone potentials encompass the Kepler $\propto -1/r$ and harmonic potential $\propto r^2$, along with many other. In this article, we revisit the classical problem of motion in isochrone potentials, from the point of view of Hamiltonian mechanics. First, we use a particularly well-suited set of action-angle coordinates to solve the dynamics, showing that the well-known Kepler equation and eccentric anomaly parametrisation are valid for any isochrone orbit (and not just Keplerian ellipses). Second, by using the powerful machinery of Birkhoff normal forms, we provide a self-consistent proof of the isochrone theorem, that relates isochrone potentials to parabolae in the plane, which is the basis of all literature on the subject. Along the way, we show how some fundamental results of celestial mechanics such as the Bertrand theorem and Kepler's third law are naturally encoded in the formalism.

\vspace{1cm}

\end{abstract}

\date{\today}


\maketitle



\section*{Introduction}

The modern theory of Hamiltonian dynamical systems was pioneered by Henri Poincaré in his celebrated \textit{New Methods of Celestial Mechanics}\footnote{to which the title of this article humbly pays tribute.} \cite{Poin}. In that work and the following ``Mémoires''\cite{Ch.12}, Poincaré proposed and explored the revolutionary idea of using geometrical and topological techniques to determine the qualitative behaviour and global properties of solutions to differential systems, rupturing with the ancestral methods devised to find exact solutions. His ideas were then developed during all the twentieth century, with applications that have proved useful to solve problems well beyond the scope of mathematics and theoretical physics \cite{Arn}. Perhaps the most ambitious problem that Poincare's methods was able to tackle was the qualitative resolution of the classical $N$-body problem, which ultimately gave birth to KAM-theory \cite{Se.03} and the stability analysis of quasi-integrable Hamiltonian systems.\\
    
At the chore of Hamiltonian mechanics lies the notion of periodic solutions. The two classical examples are the two-body (or Keplerian) problem, and the harmonic oscillator. The former is at the basis of all celestial mechanics, and the latter encodes the periodic nature of basically all integrable Hamiltonian systems. There is, however, another important feature that these two fundamental problems have in common: they are \textit{isochrone} systems, in the sense that \textit{all periodic orbits in the Kepler and harmonic potentials have a period that is only a function of the energy} of the system. In particular, this period is independent of the angular momentum of the particles orbiting these potentials, whence the name \textit{iso}chrone. This particular notion of isochrony was introduced in galactic dynamics by Michel Hénon \cite{HeI.59}, in a very specific context: finding a potential that could account for the harmonic, central regions of galaxies and their Keplerian outskirts. In his study, Hénon found a third isochrone potential, now called \textit{the} isochrone model \cite{BiTr}.\\
    
In a recent series of papers \cite{SPD,RP.20}, new light was shed on these isochrone potentials and orbits therein. In this paper, we propose to continue (and somewhat complete) this program, by examining isochrone potentials and isochrone orbits within the realm of Hamiltonian mechanics. One of our goal is to bring to light the central role and universal property of isochrone potentials, in place (and as a generalisation) of the well-established Kepler and harmonic. As a matter of fact, fundamental properties and symmetries of these two academic potentials, for example Kepler's laws of motion, the Bertrand Theorem, the Kepler Equation,  eccentric orbital elements and other fundamentals of classical gravitational mechanics, are all but special cases of the central properties of isochrone systems. In particular, they can all be understood and derived from geometrical reasoning. In \cite{SPD,RP.20} it was shown already that these results follow from \textit{Euclidean} geometry, as the set of isochrone potentials is in a one-to-one correspondence with the set of parabolae. In this article, we show that isochrony can be naturally explained and understood in the context of \textit{symplectic} geometry, the isochrony of a system being encoded into the Birkhoff invariants of its Hamiltonian formulation. To present these ideas, we have organised this paper in four sections, as follows: \\

\textbullet \, In section \ref{sec:un}, we provide a brief reminder of the general theory of test particles in radial potentials (\ref{sec:gen}), as well as a quick introduction and summary of classical results on isochrone potentials (\ref{sec:gauge}) and isochrone parabolae (\ref{sec:one}). These reminders all follow from \cite{SPD,RP.20}.\\

\textbullet \, The aim of section \ref{sec:deux} is twofold: first, we solve the general problem of isochrone dynamics in the context of Hamiltonian mechanics, by constructing a well-suited set of action-angle variables (\ref{sec:hamiso}). Then, we show that the Kepler equation of celestial mechanics, reLating the orbital time to the eccentric anomaly, actually holds for the whole class of isochrone orbits (\ref{sec:ecc}). Lastly, we use this generalised Kepler equation to derive a parametric solution for the orbital polar coordinates, in terms of a generalised eccentric anomaly (\ref{sec:para}).\\

\textbullet \, Sections \ref{sec:trois} and \ref{sec:quatre} are rather independent from the first two, and dedicated to the proof of three fundamental results within the theory of isochrone potentials. The main tool we use is the Birkhoff normal form (and Birkhoff invariants), which arises in the context of Hamiltonian mechanics. In section \ref{sec:trois}, we briefly introduce (\ref{sec:Binv}) and construct this normal form and these invariants (\ref{sec:Biso} and \ref{sec:Bgen}). We use them in section \ref{sec:quatre} to prove the fundamental theorem of isochrony (\ref{sec:fti}), the Bertrand theorem (\ref{sec:Ber}) and a generalisation of Kepler's third law to all isochrone potentials (\ref{sec:K3}). \\

For convenience, a summary of the notations used in this paper and in the previous ones \cite{SPD,RP.20} is provided in Table \ref{Table}. Several appendices contain mathematical details as well as secondary remarks worthy of interest, but that would otherwise break the natural flow of the arguments. 

\section{Isochrone potentials, isochrone orbits} \label{sec:un}

\subsection{Generalities} \label{sec:gen}

Consider, in the 3-dimensional Euclidean space of classical mechanics, a spherically symmetric body of mass density $\rho(r)$, with $r$ a radial coordinate. This body generates a radial potential $\psi(r)$ through the Poisson equation $\Delta \psi=4\pi G \rho$. We are interested in the motion of a test, unit mass particle orbiting in this radial potential. As is well-known, the spherical symmetry implies that the motion is confined in a plane orthogonal to the angular momentum vector $\vec{L}=\vec{r}\times\vec{v}$. In particular, the norm $\Lambda:=|\vec{L}|$ of the latter is conserved, as is as the mechanical energy $\xi=\tfrac{1}{2} |\vec{v}|^2+\psi(r)$. In general, i.e., for a generic potential $\psi(r)$, and when viewed in the 2-dimensional orbital plane, the quantities $(\xi,\Lambda)$ are the only two constants of motion. 

If we consider polar coordinates $(r,\theta)$ on the orbital plane, the explicit formulae for $\xi,\Lambda$ can be turned into two ordinary differential equations with respect to time for the coordinate position $(r(t),\theta(t))$ of the particle. These read
\beq \label{eom}
   \frac{1}{2} \biggl(\frac{\ud r}{\ud t} \biggr)^2 = \xi - \frac{\Lambda^2}{2r^2} - \psi(r)\,, \quad \frac{\ud \theta}{\ud t} = \frac{\Lambda}{r^2} \, .
\eeq
If the motion is bounded, then the function $r(t)$ solving this system must be periodic. Let us call \textit{radial period}, the smallest value $T\in\RR_+$ such that $r(t+T)=r(t)$ for all $t\geq 0$. The minimum (resp. maximum) values $r_p$ (resp. $r_a$) of the function $r(t)$ then corresponds, physically, to the radius at periastron (resp. apoastron). They can be obtained in terms of $(\xi,\Lambda)$ by solving the algebraic equation obtained by setting $\dot{r}=0$ in \eqref{eom}. Without loss of generality, we assume that, at initial time $t=0$, the particle is at periastron and the polar angle is $0$, so that $(r(0),\theta(0))=(r_p,0)$. The initial conditions for the coordinate velocities $(\dot{r}(0),\dot{\theta}(0))$ are then uniquely specified once $\xi,\Lambda$ are fixed, using equations \eqref{eom}.

An explicit formula can be obtained for $T$ by isolating a time element $\ud t$ from equation \eqref{eom} and integrating it over one radial period, giving 
\beq \label{defT}
   T(\xi,\Lambda) := \sqrt{2}\int_{r_p}^{r_a} \biggl( \xi-\frac{\Lambda^2}{2r^2} - \psi(r)\biggr)^{-1/2} \ud r \, .
\eeq
With the radial period $T$, another quantity of interest is the \textit{apsidal angle} $\Theta$, defined as the variation of polar angle $\theta$ during one radial period, namely $\Theta := \theta(t+T) - \theta(t)$. From equation \eqref{eom}, $\Theta$ is a constant. Integrating the equation $\dot{\theta}(t)=\Lambda/r^2$ over one radial period then gives an explicit, integral expression for $\Theta$:
\beq \label{defTheta}
   \Theta(\xi,\Lambda) := \sqrt{2}\Lambda \int_{r_p}^{r_a} \biggl( \xi-\frac{\Lambda^2}{2r^2} - \psi(r)\biggr)^{-1/2} \frac{\ud r}{r^2} \, .
\eeq
In equations \eqref{defT} and \eqref{defTheta}, both quantities $T$ and $\Theta$ have a functional dependence on the potential $\psi$, and also depend on both constants of motion $(\xi,\Lambda)$. Now we are ready to state what makes a potential $\psi(r)$ isochrone.  

\subsection{Isochrone potentials}\label{sec:gauge}

A potential $\psi$ being fixed, both $T$ and $\Theta$ now depend on the two constants of motion $(\xi,\Lambda)$. A radial potential $\psi(r)$ is said to be \textit{isochrone} when all bounded orbits within this potential are such that $T$ is independent of the angular momentum $\Lambda$: 
\beq \label{defiso}
    \psi(r) \text{ is isochrone } \Leftrightarrow \text{ } T(\xi,\Lambda) \text{ is independent of } \Lambda.
\eeq

Fundamentally, the definition of isochrony is thus encoded in the radial period $T$. However, as noticed initially in \cite{SPD}, the isochrone property of $\psi$ can be equivalently encoded in the angular part of the dynamics, via a condition on the apsidal angle $\Theta$. Indeed, we will crucially rely on the following, equivalent characterisation:
\beq \label{defisoT}
    \psi(r) \text{ is isochrone } \Leftrightarrow \text{ } \Theta(\xi,\Lambda) \text{ is independent of } \xi.
\eeq
Notice the duality between \eqref{defiso} and \eqref{defisoT}. The equivalence between the two directly follows from considerations about a special set of action-angle variables, which we will detail in section \ref{sec:hamiso}. 

Regarding both academic purposes and physical applications, the two most important radial potentials are without doubt the Kepler potential $\psi_{\rm{Ke}}$ and the harmonic potential $\psi_{\rm{Ha}}$, defined by
\beq \label{KeHa}
\psi_{\rm{Ke}}(r)=-\frac{\mu}{r} \quad \text{and} \quad \psi_{\rm{Ha}}(r)=\frac{1}{8}\omega^2 r^2 \,,
\eeq
with $\mu$ and $\omega$ two constants characterising the mass sourcing these potentials. These two potentials \textit{are} isochrone. This can be readily seen from the radial period of a particle orbiting in these potentials. They read, respectively,
\beq \label{periods}
T_{\rm{Ke}} = \frac{2\pi\mu}{(-2\xi)^{3/2}}  \quad \text{and} \quad T_{\rm{Ha}} = \frac{2\pi}{\omega} \,.
\eeq
The first equation in \eqref{periods} is Kepler's celebrated third law of motion, and the second explains our choice of normalisation for the harmonic potential (so that $\omega$ coincides with the radial angular frequency). Physically, any orbit in either of these two potentials are perfectly closed ellipses. This remarkable property holds for and only for these two potentials, a result known as Bertrand's theorem \cite{Arn,BiTr}. We shall come back to this and provide a proof of this fundamental theorem within the context of isochrony, in section \ref{sec:quatre}. As a consequence of this, it is not surprising that the apsidal angle for these two potentials is a rational multiple of $\pi$, which reads explicitly:
\beq \label{Thetas}
\Theta_{\rm{Ke}} = 2\pi \,, \quad \text{and} \quad \Theta_{\rm{Ha}} = \pi \,.
\eeq
As we can see, the isochrone property $T=T(\xi)$ and $\Theta=\Theta(\Lambda)$ are verified throughout equations \eqref{periods} and \eqref{Thetas}, indeed demonstrating the isochrone character of the Kepler and harmonic potentials.

There exists, however, other isochrone potentials. The most well-known is probably the one discovered by Michel Hénon \cite{HeI.59,HeII.59} which is usually called \textit{the} isochrone. It is given by
\beq\label{He}
\psi_{\rm{He}}(r)=-\frac{\mu}{\beta+\sqrt{\beta^2+r^2}} \,,
\eeq
where $\beta$ is a positive constant. Notice that when $\beta=0$, $\psiHe$ reduces the Kepler potential $\psiKe$. It is preferable to refer to \eqref{He} as the \textit{Hénon} potential, reserving the qualifier \textit{isochrone} for any potential with the defining property ``$T$ independent of $\Lambda$''. Two other classes of isochrone potentials exist, called the $Bounded$ potentials and the $Hollowed$ potentials. They were put forward and discussed in \cite{SPD} and \cite{RP.20}, respectively. They are given by
\beq\label{BoHo}
\psi_{\rm{Bo}}(r)=\frac{\mu}{\beta+\sqrt{\beta^2-r^2}} \,, \quad  \text{and} \quad \psi_{\rm{Ho}} = -\frac{\mu}{r^2}\sqrt{r^2-\beta^2} \,.
\eeq
The most important feature of these potentials is that, contrary to $\psi_{\rm{Ha}}$ and $\psi_{\rm{He}}$, they are not defined for all $r\in\RR_+$. Indeed, $\psiBo(r)$ is only defined for $0\leq r\leq\beta$, whence the name \textit{Bounded} potential. Whatever the initial conditions of motion, a Bounded potential confines the motion of the particle within the 3D ball defined by $r\leq \beta$, and could thus provide an effective, toy-model for physically confined systems such as quarks in baryons \cite{Mu.93}.
On the other hand, $\psiHo(r)$ is defined only when $r\geq\beta$  
and is thus, in a sense, complementary to the Bounded class $\psiBo$: where a particle in the Bounded potential cannot cross the $r=\beta$ sphere from within, a particle in the Hollowed potential $\psi_{\text{Ho}}$ cannot cross it from the outside. This class of potential could therefore be used to model classical, self-gravitating systems with central singularities, like dark matter halos \cite{Me.06}.\\


Finally, let us mention that it is possible to add to any isochrone potential a term of the form $\varepsilon+\frac{\lambda}{2r^2}$ where $(\varepsilon,\lambda)\in\RR^2$. In other words, if $\psi(r)$ is isochrone, then $\psi(r)+\varepsilon+\frac{\lambda}{2r^2}$ is also isochrone. This additional term can always be interpreted as a shift in angular momentum and energy of the particle, and has therefore been coined an $(\varepsilon,\lambda)$-gauge \cite{SPD}. This classification of isochrone potentials into four families $(\psiHa,\psiHe,\psiBo,\psiHo)$ up to a gauge-term was presented thoroughly in \cite{SPD} and enjoys a remarkable group structure. We will not make additional comments on it in the remaining of the article. In particular, the formalism used here (explained in the next section) will encompass all isochrone potentials at once, and we will not need to distinguish between the different classes, or the presence of a gauge.



\subsection{Isochrone parabolae}\label{sec:one}

What do all isochrone potentials have in common, besides the $T=T(\xi)$ property that defines them? The answer is that they are characterised by a central geometric property when viewed in other variables than $r$. This fundamental result was covered in \cite{SPD,RP.20}, but for the sake of consistency and to introduce our notations, we review it briefly in this section. \\

\begin{figure}[!htbp]
\centering
	\includegraphics[width=.65\linewidth]{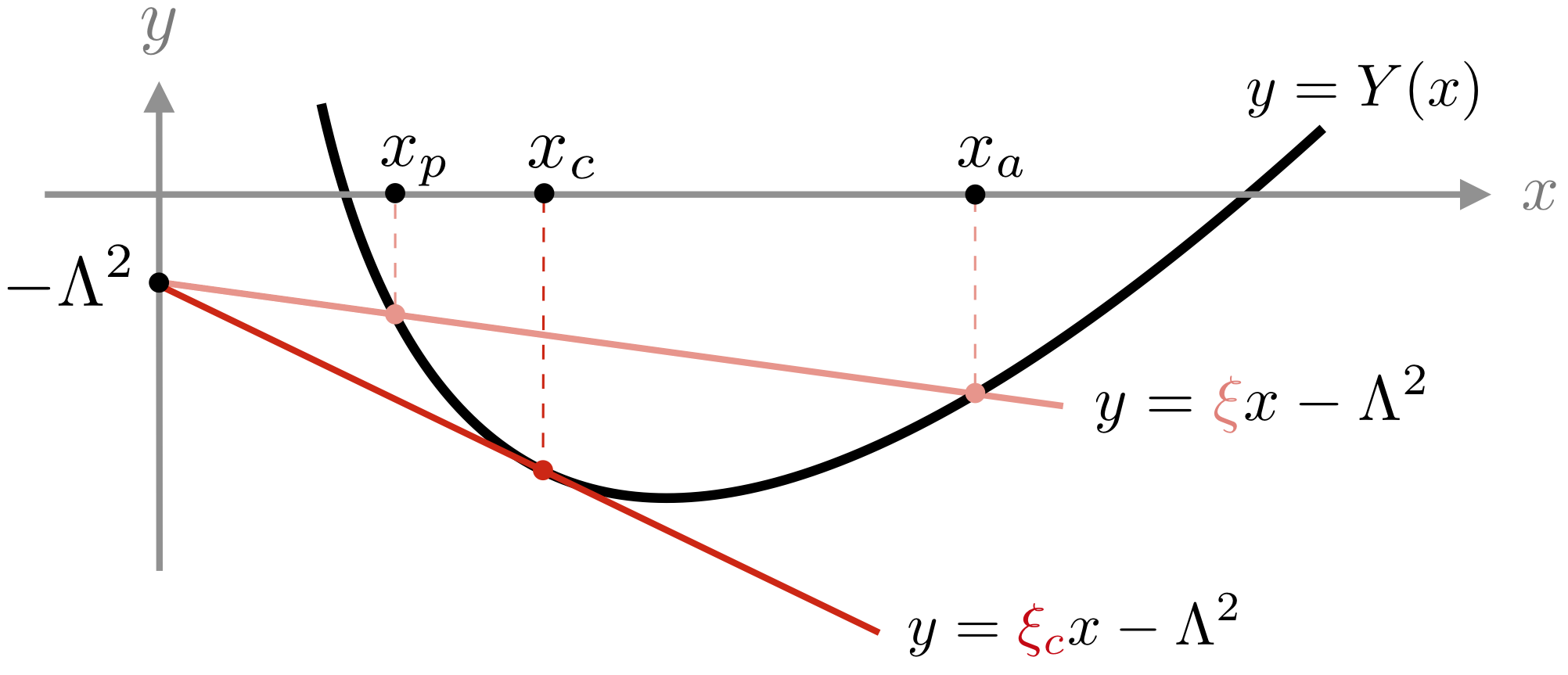}
	\caption{Construction of two orbits in the $(x,y)$-plane. The periastron and apoastron $r_p$ and $r_a$ of the physical orbit are such that $x_p=2r_p^2$ and $x_a=2r_a^2$, given by the intersections between the curve $y=Y(x)$ (potential) and the line $y=\xi x-\Lambda^2$ (particle). For a given $\Lambda$, there is a value $\xi_c(\Lambda)$ such that the orbit is circular, of radius $r_c$ such that $x_c=2r_c^2$. \label{fig:Henon}}
\end{figure}

Introducing the so-called Hénon variable $x=2r^2$ and the potential $Y(x)$ via $Y(2r^2)=2r^2\psi(r)$, the radial equation of motion \eqref{eom} can be rewritten as
\beq \label{eomx}
\frac{1}{16} \left(\frac{\ud x}{\ud t}\right)^2 = \xi x-\Lambda^2 - Y(x) \,.
\eeq
The radial motion $r(t)$ of an orbit corresponds to a solution $x(t)$ of equation \eqref{eomx}, through $x(t)=2r(t)^2$. As readily seen on equation \eqref{eomx}, the condition $\dot{x}=0$ corresponds to solutions $x$ of the algebraic equation $\xi x-\Lambda^2=Y(x)$, i.e., to intersections between the line $y=\xi x-\Lambda^2$ and the curve $y=Y(x)$, in the $(x,y)$-plane. These turning points are nothing but the periastron $x_p:=2r_p^2$ and apoastron $x_a:=2r_a^2$ of the orbit, in the $x$ variable. When there is only one intersection, at some abscissa $x_c=2r_c^2$, the associated orbit is circular. From this point of view, the variables $(x,Y(x))$ are particularly useful compared to $(r,\psi(r))$ as it allows to make a one-to-one geometric correspondence between a particle, described by $(\xi,\Lambda)$, and a given potential $\psi(r)$, described by $Y(x)$. This procedure allows to construct orbits very easily in the $(x,y)$-plane, as depicted on figure \ref{fig:Henon}.

In this article, one of the main goals is to prove the \textit{fundamental theorem of isochrony}, which lies at the core of each and every results regarding isochrone potentials \cite{SPD,RP.20}. In terms of Hénon variable $x=2r^2$ and the potential $Y(x)=x\psi(r(x))$ variable, it is simply stated as:
\beq \label{thm}
    \psi(r) \text{ is isochrone } \Leftrightarrow \text{ } Y(x) \text{ is a convex arc of parabola.}
\eeq
Although this fundamental result has been somewhat proven by Michel Hénon in \cite{HeI.59}, it is only in \cite{SPD} that a rigorous proof was provided, using complex analysis. In \cite{RP.20}, it was linked to a intrinsic geometric property verified by parabolae that dates back to Archimedes. These proofs may be somewhat unsatisfying since they need tools outside of the scope of classical mechanics. It is our aim in sections \ref{sec:trois} and \ref{sec:quatre} to show that the result \eqref{thm} is actually naturally encoded in the Hamiltonian formulation of the problem. As a consistency check, one may readily verify that the Kepler and Harmonic potentials \eqref{KeHa} are expressed in the $x$ variable as:
\beq \label{KeHax}
Y_{\rm{Ke}}(x)=-\mu\sqrt{2x} \,,\quad \text{and} \quad Y_{\rm{Ha}}(x)=\frac{1}{16}\omega^2 x^2 \,.
\eeq
Both curves in \eqref{KeHax} are indeed, arcs of parabolae. The fundamental result \eqref{thm} allows an easy and complete classification of isochrone potentials simply by classifying the parabolae in the plane. This was developed thoroughly in \cite{RP.20}, we review it briefly. 

The general definition for a parabola in the $(x,y)$ plane is an implicit, second-order equation of the form
\beq \label{parabolaimplicit}
(ax + by)^2 + cx + dy + e=0 \, ,
\eeq
with coefficients $(a,b,c,d,e)\in\RR^5$. For any parabola, these five coefficients can always be chosen such that the discriminant $\delta:=ad-bc$ of the parabola is strictly positive. When $b=0$, solving equation \eqref{parabolaimplicit} for $y$ gives a quadratic polynomial in $x$, the convex part\footnote{Isochrone potentials must correspond to a convex arc of parabola since $y=\xi x-\Lambda^2$ intersects $y=Y(x)$ twice (this is therefore a chord of $Y(x)$), and from \eqref{eomx} it follows that $\xi x-\Lambda^2\geq Y(x)$, i.e., the chord is above the curve, whence the convexity.} of which reads 
\beq \label{har}
Y(x)=-\frac{c}{d} x - \frac{e}{d} - \frac{a^2}{d}\, x^2 \,.
\eeq
The affine part (first two terms on the right) in equation \eqref{har} corresponds to the addition of a constant and a centrifugal-like term in the potential $\psi(r)$, i.e., to the gauge mentioned at the end of section \ref{sec:gauge}. The quadratic term corresponds to the usual harmonic potential, as in \eqref{KeHax}. The class of upright parabolae ($d < 0$ in \eqref{har}) corresponds to the harmonic class of isochrone potentials, all defined up to an affine term (in $x$) or a gauge (in $r$) (recall the second paragraph below \eqref{BoHo}).

\begin{figure}[!htbp]
\centering
	\includegraphics[width=.85\linewidth]{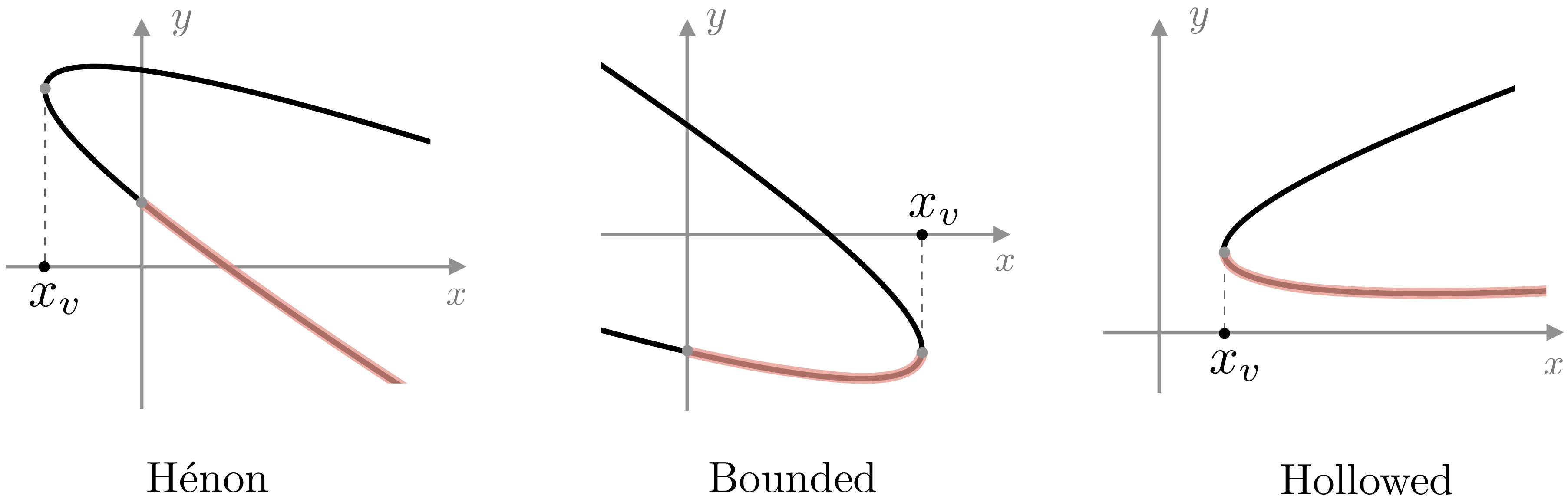}
	\caption{Three classes of (non-harmonic) isochrone parabolae. Left-orientation $(b<0)$ corresponds the Bounded class. Right-orientation $b>0$ corresponds to the H\'{e}non class ($x_v<0$) or the Hollowed class $x_v>0$. For each parabola, the physical part (i.e., where the associated $\psi(r)$ is well-defined) is highlighted in red. \label{fig:parabolas}}
\end{figure}

When $b\neq 0$, the same procedure, namely solving equation \eqref{parabolaimplicit} for $y$ and keeping the convex part, yields
\beq \label{otr}
Y(x)=-\frac{a}{b} x - \frac{d}{2b^2} - \frac{\sqrt{b\delta(x-x_v)}}{b^2}\,, \quad x_v:=\frac{4b^2e-d^2}{4b\delta} \,.
\eeq
Since $\delta>0$ and the inside of the square root must be positive, the different combination of signs for $(b,x_v)$ determines three classes of parabolae: $(b>0,x_v<0)$, $(b<0,x_v>0)$ and $(b>0,x_v>0)$ for the Hénon, Bounded and Hollowed class of parabolae, respectively. In particular, for each of these three cases, we can combine the definition $Y(2r^2)=2r^2\psi(r)$ with equation \eqref{otr} to find the three classes of potentials mentioned in \eqref{He} and \eqref{BoHo}; the parameters $\mu,\beta$ appearing there being given explicitly in terms of the Latin parameters $(a,b,c,d,e)$. For example, the Kepler potential $\psi_{\text{Ke}}$ corresponds to
\beq \label{Keplimit}
\phantom{i\quad\quad\quad} (a,b,c,d,e) \,\, = \,\, (0,1,-2\mu^2,0,0) \,, \quad \quad \text{(Kepler)}
\eeq
For the other classes of potentials, the relation between the Greek parameters $(\omega,\mu,\beta)$ of \eqref{He},\eqref{BoHo} and the Latin ones $(a,b,c,d,e)$ may be found, e.g., in section III.B.\textit{3} of \cite{RP.20}. A complete summary of the different types of isochrone potentials can be found in \cite{RP.20} (see in particular figure 7 there.) Geometrically, the sign of $b$ determines whether the parabola opens right or left, and $x_v$ is the abscissa of the point where the tangent to the parabola is vertical, as summarised on figure \ref{fig:parabolas}. In passing, we note again that each parabola in \eqref{otr} is defined up to an affine term, much like the harmonic class \eqref{har}.
In the remaining of the paper (section \ref{sec:deux} to \ref{sec:quatre}), we shall exclusively be using the Latin parameters $(a,b,c,d,e)$ that define a parabola implicitly \eqref{parabolaimplicit}, so as to state our results for any isochrone potential. However, since they are very well-known and easily derived, we will relegate results about the harmonic class \eqref{har} in appendix \ref{app:KeHo} and only consider the three other families of isochrone (with $b\neq 0$) given in equation \eqref{otr}. \\ 

We end this first section with a summary. Isochrone potentials are radial potentials in which test particles orbit with a radial period $T$ independent of its angular momentum $\Lambda$. Any (and every) isochrone potential is of the form $r\mapsto \epsilon + \frac{\lambda}{2r^2} + \psi(r)$, where $(\epsilon,\lambda)\in\RR^2$ and $\psi$ belongs to one of the four classes $(\psiHa,\psiHe,\psiBo,\psiHo)$, given in the above equations. The Harmonic class $\psiHa$ depends on one parameter $\omega\in\RR$, whereas the three other classes $(\psiHe,\psiBo,\psiHo)$ depend on two $(\mu,\beta)\in\RR_+^2$. They all have in common the property that the curve $y=Y(x)$ describes a parabola in the $(x,y)$-plane, where $x=2r^2$ is called the Hénon variable and $Y(x)=x\psi(r(x))$. This fundamental result (isochrone $\Leftrightarrow$ parabola) is referred to as the \textit{fundamental theorem of isochrony}, a proof of which will be given in section \ref{sec:quatre}. Therefore, all isochrone potentials can be represented by an implicit, 5-parameter curve \eqref{parabolaimplicit} depicting a parabola in the $(x,y)$-plane. These are the Latin parameters $(a,b,c,d,e)$ that will be used in the next sections to solve analytically the problem of motion in each and every isochrone potential.

\section{Hamiltonian solutions to the problem of motion} \label{sec:deux}

In the gravitational two-body problem of classical mechanics (see chapter 2 in \cite{BoPuV1} for a nice exposition), the orbit is a perfect ellipse. Therefore, an explicit, analytic polar equation $r(\theta)$ can be found. However, no analytic solution can be found in the form $(r(t),\theta(t))$, where $t$ is the time. However, it is possible to find a parametric solution for all three, namely $(r(E),\theta(E),t(E))$, in terms of the so-called eccentric anomaly $E$ (these classical results are recalled in appendix \ref{app:KeHo}). In this section, we derive a series of formulae that are closely related to this parametric solution of the Keplerian problem, but that is actually true of any isochrone orbit (meaning any orbit in any isochrone potential). Quite remarkably, all these formulae can be derived analytically in terms of (1) the properties of the particle $(\xi,\Lambda)$ and (2) the properties of the isochrone potential $(a,b,c,d,e)$. We derive these formulae, and compare them to the Keplerian case to motivate generalised definitions. It should be noted that some of the following results were proposed in slightly different forms as "useful formula for numerical methods" in appendix A of \cite{McGBi.90}, and in section 5.3 of \cite{BoPuV1}. In both cases, this concerns only the (non-gauged) Hénon potential, and not the whole class of isochrone.

\subsection{Hamiltonian and action-angle variables}\label{sec:hamiso}

From now on, we consider the isochrone problem from the point of view of Hamiltonian mechanics. But first, let us be more general and let $H$ be the Hamiltonian of the system made of a particle in a generic radial potential $\psi(r)$, not necessarily isochrone. In terms of the polar coordinates adapted to the orbital plane $(r,\theta)$, the canonical momenta $(p_r,p_\theta)$ simply read $(\dot{r},\Lambda)$, as is is well-known (we set the mass of the particle to $1$). The constancy of the angular momentum $\Lambda$ then follows from the fact that $\theta$ is a cyclic variable. Indeed, in these variables the Hamiltonian reads
\beq \label{H}
H(r,\theta,p_r,p_\theta)=\frac{1}{2} \biggl( p_r^2 + \frac{p_\theta^2}{r^2}\biggr) + \psi(r) \, ,
\eeq
Now we consider the problem in terms of action-angle variables. Actions may be constructed in a systematic manner by using the Poincaré invariants $J_i:=\tfrac{1}{2\pi}\oint p_i \ud q_i$, where $i\in\{r,\theta\}$ and the integral is performed over any closed curve in phase space followed during one orbital transfer (e.g., from one periastron to the following) \cite{BiTr,Arn}. For the angular part, this is almost tautological: $J_\theta:=\tfrac{1}{2\pi}\oint p_\theta \ud \theta = \Lambda$, which is a constant of motion. For the radial part, a quick computation provides 
\beq \label{j}
J_r:=\frac{1}{2\pi} \oint p_r \ud r \quad \Rightarrow \quad J_r=\frac{\sqrt{2}}{\pi} \int_{r_p}^{r_a} \biggl( \xi-\frac{\Lambda^2}{2r^2} - \psi(r)\biggr)^{1/2} \ud r \,, 
\eeq
where $r_p$ and $r_a$ are the periastron and apoastron radii, respectively, and to get the second identity we simply integrated $p_r=\dot{r}$ as given in \eqref{eom}. We now have a set of actions which we will denote $(J_r,J_\theta)=(J,\Lambda)$ from now on, for simplicity and without risk of confusion.
By definition of action-angle variables, the Hamiltonian $H$ of the system is independent of the angles. We will now show that, under the assumption of isochrony, an explicit expression for $H=H(J,\Lambda)$ can be obtained.\\

First, notice that, as provided in \eqref{j}, the radial action $J$ generally depends on both constants of motion $(\xi,\Lambda)$. Taking the partial derivatives of the rightmost equation in \eqref{j}, and comparing the result with the definitions \eqref{defT} and \eqref{defTheta} reveals\footnote{Formulae \eqref{derJ} are true in general, and explicit formulae such as the r.h.s of \eqref{j} is not necessary to derive \eqref{derJ} from $J=\tfrac{1}{2\pi}\oint p_r \ud r$. Fundamentally, this can be understood from the fact that, locally around the equilibrium (circular orbit), the pair $(H,t)$ itself defines symplectic coordinates (see \cite{Fe.13} for more details).} that \cite{SPD}
\beq \label{derJ}
\frac{T}{2\pi} = \frac{\partial J}{\partial \xi} \quad \text{and} \quad \frac{\Theta}{2\pi} = - \frac{\partial J}{\partial \Lambda} \, .
\eeq
The identities \eqref{derJ} are true of any radial potential, not necessarily isochrone. However, in the case of isochrony, by definition \eqref{defiso} we have $\partial_\Lambda T=0$, but \eqref{derJ} implies that $\partial_{\Lambda}T =- \partial_{\xi}\Theta$ (by swapping the order of derivatives using Schwartz's theorem.). Therefore, we see that $\partial_\Lambda T=0\Rightarrow\partial_{\xi}\Theta=0$ and conversely, thus recovering the equivalence between \eqref{defiso} and \eqref{defisoT}, mentioned in section \ref{sec:one}.

From now on, we assume that the potential is isochrone. Consequently,we have $T=T(\xi)$ and $\Theta=\Theta(\Lambda)$. Therefore, the PDE's in \eqref{derJ} are easily integrated and combined to give 
\beq \label{RA}
J(\xi,\Lambda) = \frac{1}{2\pi}\int \!T(\xi)\ud \xi - \frac{1}{2\pi}\int \!\Theta(\Lambda)\ud \Lambda \,,
\eeq
where any antiderivative can be considered at this stage. We emphasise that whereas \eqref{derJ} holds for any $\psi$, equation \eqref{RA} only holds for isochrone $\psi$. To make more progress towards the explicit expression of the isochrone Hamiltonian, we need to refer to \cite{RP.20} where it was shown (based on geometric arguments on parabolae) that for \textit{any} particle of energy and angular momentum $(\xi,\Lambda)$ orbiting in \textit{any} isochrone potential, parametrized by $(a,b,c,d,e)$ as explained in section \ref{sec:one}, the radial period $T$ admits a closed-form expression, given by
\beq \label{T}
T^2 = - \frac{\pi^2}{4} \frac{\delta}{(a+b\xi)^3} \,,
\eeq
where we recall that $\delta:=ad-bc>0$. Of course, $T$ as given by equation \eqref{T} does not depend on the angular momentum $\Lambda$ of the particle, by definition of an isochrone potential (cf \eqref{defiso}). Also derived in \cite{RP.20} was a similar formula for the apsidal angle $\Theta$, which reads
\beq \label{Theta}
 \frac{\Theta^2}{\pi^2\Lambda^2} = \frac{2b^2\Lambda^2-d}{b^2\Lambda^4-d\Lambda^2+e} + \frac{2b}{\sqrt{b^2\Lambda^4-d\Lambda^2+e}} \,,
\eeq
and which is indeed independent of the energy $\xi$ (cf \eqref{defisoT}). With the help of formulae \eqref{T} and \eqref{Theta}, we can integrate explicitly\footnote{Although \eqref{T} and \eqref{Theta} hold for any isochrone, including the harmonic class; starting from equation \eqref{action}, most expressions differ in the harmonic case (cf appendix \ref{app:KeHo}), because of the condition $b=0$.} equation in \eqref{RA} and obtain
\beq \label{action}
J(\xi,\Lambda) = \frac{1}{2b} \sqrt{\frac{-\delta}{a+b\xi}} - \frac{R(\Lambda)}{2b} \,,
\eeq
where for convenience we introduced the function $R(\Lambda)$ independent of $\xi$ and given by 
\beq \label{R}
R(\Lambda) := \sqrt{2b^2\Lambda^2 - d + 2b \sqrt{b^2\Lambda^4 - d\Lambda^2 + e}} \,.
\eeq
It should be noted that while performing the integrals from \eqref{RA} to \eqref{action}, a constant of integration should be included in the latter expression. However, that constant can be shown to vanish since $J$ should reduce to the well-known \cite{BiTr} radial action $J_{\text{Ke}} = \mu/\sqrt{-2\xi}-\Lambda$ in the Kepler potential $\psi(r)=-\mu/r$ (which is isochrone), corresponding to the limit \eqref{Keplimit}. Equation \eqref{action} gives an exact formula for the radial action of all non-harmonic isochrone potentials. For the harmonic class ($b=0$), the computation is given in appendix \ref{app:KeHo} (see equation \eqref{Jhar} there).\\

\begin{figure}[!htbp]
	\includegraphics[width=0.7\linewidth]{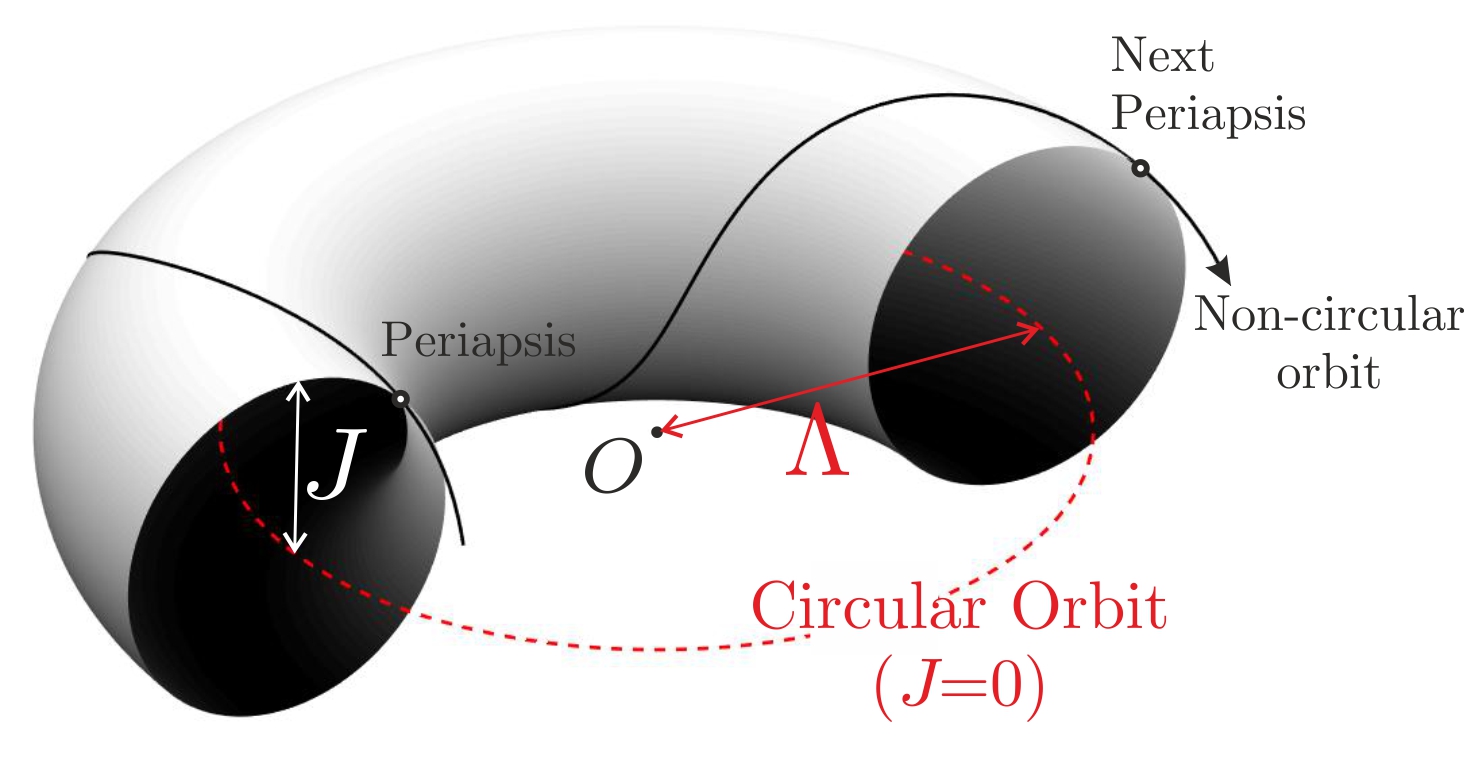}
	\caption{One torus $(J,\Lambda)$ in the phase space, depicting a generic (non-circular) orbit (black curve). The circular orbit $(J=0$ with the same angular momentum $\Lambda$ is depicted in red. \label{fig:torus}}
\end{figure}

Going back to the Hamiltonian $H(J,\Lambda)$, for any pair $(J,\Lambda)$ corresponding to a well-defined orbit, the numerical value of $H$ is actually the energy $\xi$ of the particle. Therefore, we may solve equation \eqref{action} for $\xi$ in terms of $(J,\Lambda)$, to obtain the expression of $H(J,\Lambda)$. This readily gives
\beq \label{ham}
H(J,\Lambda) = -\frac{a}{b}-\frac{\delta}{b\bigl(2bJ + R(\Lambda) \bigr)^2} \,,
\eeq
with $R(\Lambda)$ was given in \eqref{R}. Equation \eqref{ham} provides the general expression for the Hamiltonian of a particle in any non-harmonic isochrone potential in action-angle variables. (see equation \eqref{Hhar} of appendix \ref{app:KeHo} for the harmonic class). In the Keplerian limit \eqref{Keplimit}, we recover the Hamiltonian of the classical two body problem in terms of the Delaunay variables (see e.g. equation (E.1) of \cite{BiTr}). It coincides with (and generalises) the Hamiltonian of the Hénon potential as discussed in section 3.5.2 of \cite{BiTr}. In action-angle variables, the equations of motion for the isochrone orbit are in their simplest form, given by the constancy of $(J,\Lambda)$ and the linear-in-time evolution of the associated angles, namely 
\beq \label{angles}
z_J(t) = \omega_J t + z_J(0) \quad \text{and} \quad z_{\Lambda}(t) = \omega_{\Lambda} t + z_{\Lambda}(0) \,. 
\eeq
where the Hamiltonian frequencies $\omega_i$ read, by definition, 
\beq
\omega_J := \frac{\partial H}{\partial J} \quad  \text{and} \quad  \omega_\Lambda := \frac{\partial H}{\partial \Lambda} \,.
\eeq
In action-angle variables, the four-dimensional phase space can be represented by embedding a torus of radii $J,\Lambda$ in $\RR^3$, allowing for a particularly nice representation, see figure \ref{fig:torus}. However, reLating the angle variables to the polar coordinates $(r,\theta)$ remains to be done. The easiest way is to express the time $t$ that appears in \eqref{angles} in terms of $(r,\theta)$. This will enable us to derive a generalisation of the Kepler equation and Kepler's third law, as well as true/eccentric anomaly relations, which we explore in the next subsection. A straightforward computation from equation \eqref{ham} reveals that the frequencies read 
\beq \label{freq}
\omega_J = \frac{4\delta}{\bigl(2bJ+R(\Lambda)\bigl)^3} \quad \text{and} \quad \omega_\Lambda = \frac{2\delta R'(\Lambda)}{\bigr(2bJ+R(\Lambda)\bigl)^3} \,.
\eeq
Whenever the orbit is closed in real space, it should also be in phase space. Consequently, the ratio number. Indeed, computing the ratio $\omega_\Lambda/\omega_J$ using equations \eqref{freq} and comparing the result to equation \eqref{Theta} readily gives
\beq \label{ratiofreq}
\frac{\omega_\Lambda}{\omega_J} = \frac{\Theta(\Lambda)}{2\pi} \,.
\eeq

It is quite remarkable that all isochrone potentials admit an universal and closed-form expression for the Hamiltonian in action-angle variables. Coupled to the large variety of properties (recall section \ref{sec:gauge}) that these potentials offer, this allows for a remarkable pedagogical tool to teach Hamiltonian mechanics with much more applications than the usual harmonic oscillator and two-body problem. From the fundamental point of view, it is very tempting to build toy-models for different types of systems in terms of isochrone potentials, as it is rather rare to have analytic expressions available for entire systems. In fact, analyticity has probably been the main reason for the success of the Hénon potential in this context, as a generalisation of the Kepler potential with closed-form expressions. We emphasise that this characteristic (availability of closed-form formulae) holds for all isochrones.

\subsection{Kepler equation and eccentric anomaly} \label{sec:ecc}

In \cite{RP.20}, it was shown that the equations of motion for a subclass of isochrone orbits (namely those associated with parabolae crossing the origin of the $(x,y)$-plane) could be integrated analytically in a parametric expression of the type $(r(s),\theta(s))$ for some parameter $s\in\RR$. The parameter used in these expression was then a pure mathematical quantity, bearing, \textit{a priori}, no physical meaning. In particular, these parametric equations were obtained by reLating any isochrone orbit to a Keplerian one, through a linear transformation acting on their respective arcs of parabolae, in the $(x,y)$-plane. In this section, we show that these formulae can be obtained (1) by direct integration, (2) without making any assumption as to the subclass of isochrone, (3) such that the parameter $s$ admits a clear, physical interpretation. \\

We suppose that a particle of energy and angular momentum $(\xi,\Lambda)$ orbits an isochrone potential $\psi(r)$. As argued before, $\psi(r)$ is in a one-to-one correspondence with a convex arc of parabola $y=Y(x)$. We start by deriving an explicit formula for the time $t$ elapsed during orbit. Let $T$ be the radial period and $t\in [0;T/2]$ be an instant between the initial-time periastron $r(0)=r_p$ and apoastron $r(T/2)=r_a$. By isoLating $\ud t$ in the radial equation of motion \eqref{eomx} and integrating, we readily obtain 
\beq \label{temp}
t = \frac{1}{4}\int_{x_p}^{x}  \bigl(a_0 + a_1 x+\sqrt{a_2 x + a_3}\,\bigr)^{-1/2} \ud x \,,
\eeq
where, for the sake of simplicity, we temporarily introduced the following coefficients that depend on the particle $(\xi,\Lambda)$ and the potential $(a,b,c,d,e)$:
\beq \label{coefficients}
(a_0,a_1,a_2,a_3) 
=\biggl(\frac{d}{2b^2}-\Lambda^2,\xi+\frac{a}{b},\frac{\delta}{b^3},\frac{d^2-4b^2e}{4b^4}\biggr)\,.
\eeq
Equation \eqref{temp} is nothing but the integral \eqref{T} written in terms of $x=2r^2$, and for an isochrone $Y(x)$. The change of variables $u=\sqrt{a_2x +a_3}$ then turns the term in parenthesis in \eqref{temp} into a pure quadratic, namely
\beq \label{temp2}
t = \frac{\sqrt{u_0}}{\sqrt{2}a_2}\int_{u_p}^{u} \frac{u\,\ud u}{\sqrt{v_0-(u-u_0)^2}} \,,
\eeq
where $(u_0,v_0)$ are the coordinates of the apex of that quadratic, given by $u_0 = -a_2/2a_1$ and $v_0 = u_0^2 + 2a_0u_0 + a_3$. Note that in the $u$ variable, the periastron (lower bound of the integral \eqref{temp2}) corresponds to the smallest root of the quadratic in the denominator, namely $u_p=u_0-\sqrt{v_0}$. To integrate equation \eqref{temp2}, we start by turning the quadratic $v_0-(u-u_0)^2$ in canonical form by performing the linear transformation $s=(u-u_0)/\sqrt{v_0}$, so that $s=-1$ when $u=u_p$. This turns \eqref{temp2} into
\beq\label{intint}
\Omega\,t = \int_{-1}^{s} \frac{1+\epsilon s}{\sqrt{1-s^2}}\,\ud s \,, \quad \text{where }\,\,\epsilon=\frac{\sqrt{v_0}}{u_0}\,,\quad\Omega = \frac{\sqrt{2}a_2}{u_0^{3/2}} \,.
\eeq
Notice that by construction, $0<\epsilon<1$, since $u_p=u_0-\sqrt{v_0}>0$. The integral in \eqref{intint} may then be finally integrated by defining an angle $E\in[0,\pi]$ such that $s=-\cos E$, with $s=0$ at periastron $(t=0\Leftrightarrow s=-1)$. Integrating in this fashion, we obtain the sought-after, expression for $t$, which takes the form of a generalised Kepler equation
\beq \label{KE}
\Omega \, t = E -\epsilon \sin E \,.
\eeq

Equation \eqref{KE} looks exactly like the Kepler equation \eqref{Keplereq} found in the classical two-body problem.  The expression of the constants $(\Omega,\epsilon)$ can be given in terms of the generic Latin parameters $(a,b,c,d,e)$ and $(\xi,\Lambda)$ that characterise the potential and the particle, respectively. These expressions read:
\begin{subequations} \label{Omegepsi}
\begin{align}
    \Omega^2 &= -16\Delta(a+b\xi)^3 \label{Omega}\,, \\
    \epsilon^2 &= 1+2\delta^{-1}(2b^2\Lambda^2-d)(a+b\xi)+\delta^{-2}(d^2-4b^2e)(a+b\xi)^2 \,,
\end{align}
\end{subequations}
where $\delta=ad-bc$. Naturally, we may identify $\epsilon$ and $E$ as the \textit{isochrone eccentricity} and \textit{isochrone eccentric anomaly}, respectively. The isochrone eccentricity verifies $0\leq\epsilon<1$, vanishes only for circular orbits and coincides with the Keplerian eccentricity in the appropriate limit \eqref{Keplimit}. The isochrone eccentric anomaly is a well-defined angle and coincides with its Keplerian counterpart as well (as we will see in the next subsection). It should be stressed that the frequency $\Omega$ also coincides with $2\pi/T$, as we see by comparing \eqref{Omega} and \eqref{T}. In other words, the left-hand side of the generalised Kepler equation \eqref{KE} involves the frequency $\Omega$ of the radial motion $r(t)$. In particular, combining the angle coordinates \eqref{ratiofreq} and the Kepler equation \eqref{KE} provides
\beq \label{zKep}
z_J = E - \epsilon \sin E \,, \quad z_{\Lambda} = \frac{\Theta}{2\pi} (E - \epsilon \sin E) \,,
\eeq
where we have set $(z_J,z_{\Lambda})=(0,0)$ at $t=0$. It is clear from \eqref{zKep} that $z_J$, the angle variable associated to the radial action $J$, generalises in fact the Keplerian mean anomaly. 


\subsection{Parametric polar solution} \label{sec:para}

Now that an eccentric anomaly $E$ has been introduced via Kepler's equation \eqref{KE}, we derive its relation to the orbital radius $r$ (or equivalently $x=2r^2$) and the polar angle $\theta$. 

\subsubsection{Radial motion}

For the radial part $r(E)$, we may simply go through the different changes of variables used to compute the integral \eqref{temp} in the last subsection, but in reverse, i.e., $E\mapsto s\mapsto u\mapsto x$. After some easy algebra, we find that\footnote{There is a subtlety in the case $a_2<0$, since then the function $u(x)=\sqrt{c_2 x+c_3}$ is decreasing. This is resolved by keeping track of $\text{sign}(a_2)=\text{sign}(b)$, which results in the $\pm |b|$ in \eqref{xE}.}
\beq \label{xE}
x(E) = \frac{4b^2e-d^2}{4b\delta} \pm \frac{\delta}{4|b|(a+b\xi)^2}\,(1-\epsilon \cos E)^2\,.
\eeq
where $\pm$ corresponds to the sign of $b$. We note that the first term on the right-hand side is actually $x_v$, the abscissa of the point with vertical tangent on the parabola introduced in equation \eqref{otr}. Whenever the potential is Keplerian, then $x_v=0$ (and $b>0$) and we recover the classical link $r=\alpha_{\text{Ke}}(1-\epsilon \cos E)$, where $\alpha_{\text{Ke}}$ is the semi-major axis of the Keplerian ellipse. This motivates the following definition for an \textit{isochrone semi-major axis} $\alpha$ such that the orbital radius $r(E)$ reads
\beq \label{smaxis}
r(E) = \sqrt{\frac{x_v}{2} \pm  \alpha^2(1-\epsilon \cos E)^2}\,, \quad \text{where} \quad \alpha^2 := \frac{\delta}{8|b|(a+b\xi)^2} \,.
\eeq
This isochrone semi-major axis is, in general, not related to an ellipse axis, as isochrone orbits are not, in general, ellipses \cite{RP.20}. However, it coincides with its Keplerian counterpart in the proper limit, and comparing equations \eqref{smaxis} with \eqref{Omegepsi} reveals the equality 
\beq \label{third}
\Omega^2 \alpha^3 = \sqrt{\frac{\delta}{2|b|^3}}\,.
\eeq
which we recognise a the generalisation of Kepler's third law of motion, reLating the (square of) the orbital frequency to the (cube) of the semi-major axis. Indeed, the Keplerian limit \eqref{Keplimit} of equation \eqref{third} gives $\Omega^2\alpha^3=\mu$, a well-known formulation of Kepler's third law \cite{BiTr}. In fact, the quantity appearing on the right-hand side has dimension of mass and is exactly the mass parameter $\mu$ appearing in equations \eqref{He} and \eqref{BoHo}, see equation (3.12) of \cite{RP.20}.

\subsubsection{Angular motion}

For the angular motion, we adapt the strategy developed in \cite{RP.20} and first construct a differential equation of which $\theta(E)$ is a solution. We can do this with the Leibniz rule as follows
\beq \label{bup}
 \frac{ \ud \theta}{\ud E} = \frac{ \ud \theta}{\ud t}\, \frac{\ud t}{\ud E} = \frac{\Lambda}{\Omega} \frac{1-\epsilon \cos E}{r(E)^2} \,,
\eeq
where in the second equality we used the angular equation of motion $\dot{\theta}=\Lambda/r^2$ and the Kepler equation \eqref{KE}. To obtain an expression $\theta(E)$ from \eqref{bup}, we simply need to inject the expression of $r(E)$ given in \eqref{smaxis} and integrate the result. By doing so, we readily obtain
\beq \label{int0}
\theta(E) =  \frac{2\Lambda}{\Omega} \int_0^E \frac{1-\epsilon \cos \phi}{x_v + 2\alpha^2 (1-\epsilon \cos \phi)^2} \ud \phi \,,
\eeq
where $\alpha$ is given by \eqref{smaxis} and $x_v$ depends only on the potential, cf. \eqref{otr}. When $x_v\leq 0$, which corresponds to the Hénon class of potentials, the integral can be easily integrated. Indeed, a partial fraction decomposition gives
\beq \label{int}
\theta(E) = \frac{\Lambda }{2\Omega\alpha^2} \sum_{\pm} \frac{1}{1\pm\zeta} \int_0^E \frac{\ud \phi} {1 - \epsilon_{\pm} \cos \phi} \,, 
\eeq
where $\epsilon_{\pm}=\epsilon/(1\pm\zeta)$ with $\zeta^2=-x_v/2\alpha^2$ and is such that $0\leq \zeta \leq \epsilon < 1$, so that the integrals are well-defined. Equation \eqref{int} can be further simplified by computing explicitly the integral, and thus provides the final formula for $\theta$ in terms of $E$, namely
%
%
\beq \label{thetaE}
 \theta(E) = \frac{\Lambda }{\Omega\alpha^2} \sum_{\pm} \frac{\epsilon_{\pm}}{\sqrt{1-\epsilon_{\pm}^2}} \arctan \biggl( \sqrt{\frac{1+\epsilon_{\pm}}{1-\epsilon_{\pm}}} \tan \frac{E}{2} \biggr) \,.
\eeq
The Keplerian limit of equation \eqref{thetaE} consists in taking $\zeta\propto x_v=0$ such that $\epsilon_{\pm}=\epsilon$, and thus provides a sum of two identical terms, recovering the well-known Keplerian result \eqref{elemts}. For consistency, one can check that when $E=\pi$, which should correspond to the apoastron of the orbit, equation \eqref{thetaE} coincides with the general expression \eqref{Theta} of the apsidal angle $\Theta$, which satisfies $\theta(\pi)=\Theta/2$ by definition. \\

As final word, let us mention that the derivation of formula \eqref{thetaE} relies on the crucial assumption that $x_v\leq 0$, and thus only holds for the Hénon class of isochrone potentials (recall figure \ref{fig:parabolas}). When $x_v\geq 0$, the auxiliary quantity $\zeta$, defined by $\zeta^2=-x_v/2\alpha^2$, becomes imaginary and the $\epsilon_{\pm}$ are now complex. It turns out that formula \eqref{thetaE} also holds for this case. The reason is that, although both terms $+$ and $-$ in the $\sum_{\pm}$ sum are complex numbers, they are conjugate to one another. Their sum is therefore twice their real part, and is thus a real quantity. We provide the details of this in appendix \ref{app:complex}. In particular, formula \eqref{thetaE} for imaginary $\zeta$ is mathematically well-defined and the resulting $(r(E),\theta(E))$-orbit does coincide with the true isochrone dynamics. \\

Once again, we end this section by a summary of the results. The equations of motion for a test particle in any isochrone potential can be solved analytically in the parametric form $(r(E),\theta(E))$ where $E$ is a parameter that reduces to the Keplerien eccentric anomaly. These equations are given in \eqref{smaxis} and \eqref{thetaE}. The polar coordinates $(r,\theta)$ along the orbit can also be related to orbital time $t$ through a generalisation of the Kepler equation \eqref{KE}, that holds for any isochrone orbit. Finally, we have shown that the radial action variable $J$ is particularly well-adapted to the isochrone problem, as it (1) splits into a sum of $\xi$- and $\Lambda$-dependent terms and (2) can be used to derive the general Hamiltonian of the dynamics in action-angle variables $(J,\Lambda)$. 


\section{Birkhoff normal forms and invariants} \label{sec:trois}

The fundamental theorem of isochrony \eqref{thm} is what allows to derive all the analytical results for isochrone potentials and orbits therein, as we did in section \ref{sec:deux}. This theorem was first proven by Michel Hénon in his seminal paper \cite{HeI.59}, although not without some (minor) mistakes. It was then discussed in \cite{SPD} by borrowing techniques from complex analysis, and in \cite{RP.20} using Euclidean geometry but necessitating an abstract mathematical result. In any case, at present, a self-consistent and natural proof of this central theorem relying only on classical mechanics, is nowhere to be found, to our knowledge. It is our goal, in the present and following sections to introduce and exploit a powerful tool of Hamiltonian mechanics: the \textit{Birkhoff normal form}. In a nutshell, the Birkhoff normal form allows to (quantitatively and rigorously) probe the neighbourhood of equilibrium points in phase space, to obtain information on their stability, and thus on the integrability of the underlying Hamiltonian. \\

There exists a lot of specialised literature on this topic. Yet, introductory material on normal forms may be hard to find for non-specialists. Among the most accessible, we found that Arnold's classical textbook (\cite{Arn}, appendix 7) and Hofer \& Zehnder's lectures (\cite{HoZe.12}, sections 1.7 and 1.8) are particularly relevant (see also section (8.5) of \cite{BoPuV2} and \cite{Pi.13}). Other examples of accessible expositions (with applications) may be found in \cite{BoPuV1} (for the stability of the Lagrange points), and \cite{Gr.07} (for solving PDE's). Other notable references, namely \cite{FeKa.04,Fe04,ChPi.11}, present explicit computations of normal forms and use them to study the stability of the (restricted) $N$-body problem. The latter have largely motivated the present derivation.\\

Applications of our method go beyond the sole isochrone theorem \eqref{thm}, as it allows us to prove the Bertrand theorem \cite{Arn}, as well as the generalised Kepler's third laws \eqref{T},\eqref{Theta}. We relegate these applications to section \ref{sec:quatre}, and only focus on the derivation of the normal form in the present section, which we organise as follows:
\begin{itemize}
    \item in subsection \ref{sec:Binv} we introduce the notion of Birkhoff normal form and Birkhoff invariants in a very simple case, sufficient for our purpose;\vspace{-2mm}
    \item in subsection \ref{sec:Bgen} we write the Birkhoff normal form $N_1$ for the Hamiltonian of a particle in a \textit{generic} potential, which encodes information on the the potential $Y(x)$;\vspace{-2mm}
    \item in subsection \ref{sec:Biso} we write the Birkhoff normal form $N_2$ for the Hamiltonian of a particle in an \textit{isochrone} potential, using the radial action \eqref{j} well-adapted to isochrony.\vspace{-2mm}
\end{itemize} 

\subsection{Normal form and Birkhoff invariants}\label{sec:Binv}


For the sake of simplicity we will only cover the very basics of normal forms and refer to the above literature for the details. In particular, we consider a 1-dimensional problem (2-dimensional phase space), but all can be generalised to any $2n$-dimensional phase space (see, e.g., section 1.8 of \cite{HoZe.12}). Let $H(q,p)$ be a Hamiltonian defined in terms of coordinates $(q,p)\in\RR^2$. We assume, without any loss of generality, that the origin $(q,p)=(0,0)$ is an elliptic\footnote{We focus on elliptic equilibria since we want to describe a periodic motion of the particle.} equilibrium point. A classical theorem (due initially to Birkhoff \cite{Bi.27} and then refined/generalised since then \cite{HoZe.12}) then says that \textit{there exists a local coordinate transformation $(q,p)\mapsto (\rho,\varphi)$ such that the Hamiltonian takes the form}
\beq \label{HBirk}
H(\rho,\varphi) = \mathfrak{l} + \mathfrak{b} \rho + \frac{1}{
2} \mathfrak{B} \rho^2 + o(\rho^2) \,.
\eeq
The real numbers $(\mathfrak{l}, \mathfrak{b}, \mathfrak{B})$ are called Birkhoff invariants of zeroth, first and second order, respectively\footnote{In the general case $(q,p)\in \RR^n \times \RR^n$, then $\rho\in \RR^n$ and, accordingly, the Birkhoff invariants $\mathfrak{b}$ and $\mathfrak{B}$ are linear and bilinear forms, respectively.}. They depend exclusively on $H$ (not on the mapping $(q,p)\mapsto (\rho,\varphi)$). They encode the information on the geometry of phase space around the equilibrium point. In whole generality, Birkhoff's theorem gives much stronger results than the result \eqref{HBirk}. In particular, it holds for any dimensions, explains how to extend \eqref{HBirk} to any order in the powers of $\rho$, makes a distinction between resonant or non-resonant orbits. However, quadratic order will be sufficient for our purposes, and we will refer the reader to section (1.8) of \cite{HoZe.12} (and references therein) for a more detailed and rigorous exposition. \\

\begin{figure}[!htbp]
\centering
	\includegraphics[width=.55\linewidth]{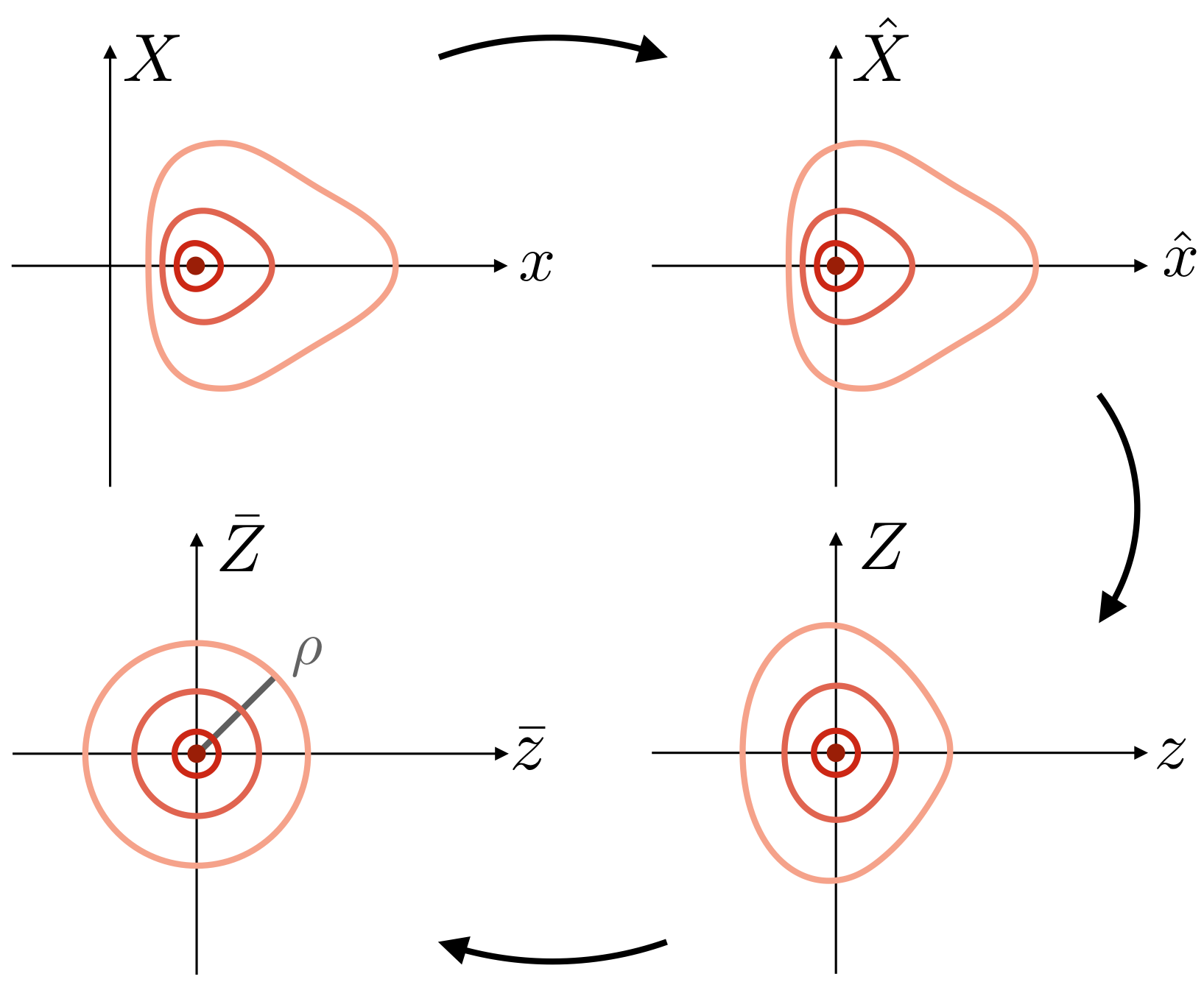}
	\caption{The circular orbit (red point) and three non-circular orbits (red curves) in the 2-dimensional phase space under each transformation. The map $(x,X)\mapsto(\hat{x},\hat{X})$ translate $x_c$ at the origin, and $(\hat{x},\hat{X})\mapsto(z,Z)$ circularises the orbits in the only in the close vicinity of the origin (the outer curves are not circular). Then $(z,Z)\mapsto(\bar{z},\bar{Z})$ circularises a larger neighbourhood of the origin (all curves circular up to $O(\rho^2)$) allowing to construct polar action-angle variables $(\rho,\varphi)$ in that region. \label{fig:PStrans}}
\end{figure}

The main feature of the variables $(\rho,\varphi)$ in \eqref{HBirk} is that, up to $o(\rho^2)$ corrections, they are action-angle coordinates, as $H$ does not depend on the angle $\varphi$. In this work, we call \textit{normal form} of $H(\rho,\varphi)$, and denote by $N(\rho)$, the quadratic part of \eqref{HBirk}, namely
\beq \label{defNorm}
N(\rho) = \mathfrak{l} + \mathfrak{b} \rho + \frac{1}{
2} \mathfrak{B} \rho^2 \,.
\eeq
Heuristically, the quantity $N(\rho)$ in \eqref{defNorm} can be seen as a Hamiltonian that is (1) completely integrable and (2) describes the same dynamics as $H$ in \eqref{HBirk} in the $O(\rho^2)$-neighbourhood of the equilibrium $(0,0)$. Owing to the unicity of the Birkhoff invariants, the normal form \eqref{defNorm} is itself unique, in the sens that a change of action-angle coordinates that leaves the equilibrium point at the origin must be the identity (see \cite{An.81} for a detailed exposition as well as appendix \ref{app:B}). A natural method to construct a normal form is crystallised in figure \ref{fig:PStrans}, which shows the successive steps one may use to transform the geometry of the phase space $(q,p)$ around $(0,0)$, so as to introduced polar-symplectic coordinates $(\rho,\varphi)$ (cf subsection \ref{sec:polar}). In fact, in section \ref{sec:Bgen} will will explicitly provide a constructive example of such $(q,p)\mapsto(\rho,\varphi)$ mapping, that brings $H(q,p)$ into normal form \eqref{HBirk}. 

\subsection{Birkhoff normal form: generic radial potential}\label{sec:Bgen}

Let us start with the Hamiltonian of a particle in a radial potential $\psi(r)$, as given by \eqref{H}, rewritten here for convenience as:
\beq \label{HH}
H(r,R,\theta,\Lambda)= \frac{R^2}{2} + \frac{\Lambda^2}{2r^2} + \psi(r) \, ,
\eeq 
where $(r,\theta)$ are the coordinates and $(R,\Lambda)$ their conjugated momenta. The complete, 4-dimensional phase space of the dynamics is a subset of $\RR_+ \times \RR \times [0;2\pi[ \times \RR_+\ni(r,R,\theta,\Lambda)$. However, since $\theta$ does not appear explicitely in \eqref{HH} and its conjugated momentum $\Lambda$ is constant, $(\theta,\Lambda)$ is already a pair of angle-action coordinates. Therefore, it can be practical to think of \eqref{HH} as a 1-dimensional family of Hamiltonians parametrised by $\Lambda$. In this way, we just need to focus the radial part $(r,R)$ of the dynamics, and perform successive symplectic transformations to reach a normal form, the coefficients of which will thus be $\Lambda$-dependent. With this \textit{2-dimensional phase space} point of view, we will write $H(r,R)$ instead of $H(r,R,\theta,\Lambda)$ to stick with the notations of section \ref{sec:Binv}, with no risk of confusion. Moreover, while performing successive symplectic changes of coordinates on the phase space $(r,R)\in\RR_+\times\RR$, we will keep the (lower case/upper case) notation for a coordinate ($r,x,z,\ldots$) and its conjugated momentum ($R,X,Z,\ldots$). \\

Although we have tried to be as pedagogical as possible (and we believe these computations are interesting in themselves), the following subsections are rather technical. On first reading (or for the reader in a hurry), it is possible to skip the following steps and just assume that there exists a pair of variables $(\rho,\varphi)$, such that the Hamiltonian \eqref{HH} admits a normal form $N_2(\rho)$, given by equation \eqref{N1} below; before directly proceeding with subsection \ref{sec:Binv}.

\subsubsection{Hénon variable and circular orbits}

The Hamiltonian \eqref{HH} describes the same system as the Hamiltonian in \eqref{HJL} only if the potential is isochrone. Since the following computations are true for any radial potential (not just isochrones) we stay general and relegate the isochrone assumption to the next subsection. Still, as we have the isochrone theorem \eqref{thm} in mind, we would prefer to speak in terms of $(x,Y(x))$ instead of $(r,\psi(r))$. Therefore, the first step is the change of variables $(r,R)\mapsto(x,X)$, where $x=2r^2$ and $X$ is the canonical momenta associated to $x$. The transformation is easily seen to be symplectic if and only if $R=\sqrt{8x}X$. In these variables, the Hamiltonian \eqref{HH} now reads 
\beq \label{Hnew}
H(x,X) = 4xX^2+\frac{\Lambda^2}{x}+\frac{Y(x)}{x} \,.
\eeq
where we recall that $Y(x)=x\psi(r(x))$. The derivation of a normal form starts with a choice of equilibrium point around which to write it. 
Using \eqref{Hnew} for a given $\Lambda$, these points are simply given by $(x,X)=(x_c,0)$, where $x_c=x_c(\Lambda)$ is a solution to the algebraic equation 
\beq \label{xc}
x_c Y'(x_c)-Y(x_c)=\Lambda^2 \,. 
\eeq
At this equilibrium, we have $X=0\rightarrow\dot{r}=0$, thus corresponding to circular orbits, the radius $r_c$ of which is such that $x_c=2r_c^2$ and $x_c$ solves \eqref{xc}. In the complete 4-dimensional phase space, there exists a family of such circular orbits, parametrised by $\Lambda$. 
%

\subsubsection{TransLating the equilibrium at the origin}

Now we are going to write the Birkhoff normal form of $H$ around a given circular orbit $(x,X)=(x_c,0)$. The main goal is to fix a $\Lambda$ and to circularise the phase space around the equilibrium $(x_c,0)$, in order to introduce symplectic polar coordinates, following the discussion in section \ref{sec:Binv}. \\

We start by transLating the equilibrium $(x,X)=(x_c,0)$ to the origin, by setting $(\hat{x},\hat{X})=(x-x_c,X)$ and then Taylor-expanding in the $\hat{x}$ variable, small by assumption. We obtain\footnote{\label{fn1}In the 4D phase space, this change of variable is rendered symplectic by changing the angle accordingly. For example, the mapping $(x,\theta,X,\Lambda) \mapsto (x-x_c(\Lambda) ,\hat{\theta},\hat{X},\Lambda)$ is symplectic if we take $\hat{\theta}=\theta-x_c'(\Lambda)X$.} the following expression
%
%
\beq \label{Hhat}
H(\hat{x},\hat{X}) = Y_1 + 4 x_c \hat{X}^2 + \frac{Y_2}{2 x_c} \hat{x}^2 + 4\hat{X}^2\hat{x} + c_3  \hat{x}^3 + c_4 \hat{x}^4 + o(\hat{x}^4) \,,
\eeq
where we introduced the convenient notation $Y_n:=Y^{(n)}(x_c)$, and defined the following coefficients that depends on the derivatives of $Y(x)$ at $x=x_c$, namely
\beq \label{coeff}
c_3 = \frac{x_c Y_3 - 3 Y_2}{6 x_c^2} \,, \quad \text{and} \quad c_4 = \frac{12 Y_2 - 4x_c Y_3 + x_c^2 Y_4}{24 x_c^3} \,.
\eeq
In these variables, the circular orbits are at the origin $(\hat{x},X)=(0,0)$, and in the 4D phase space each coefficient depends on $\Lambda$ through $x_c = x_c(\Lambda)$ (recall equation \eqref{xc}). Notice that the energy of the circular orbit is $H(0,0)=Y_1=Y'(x_c(\Lambda))$. This is in agreement with the way orbits are constructed in the Hénon variables, as explained around figure \ref{fig:Henon}. Lastly, for the sake of completeness, let us deduce from \eqref{Hhat} the nature of the equilibrium $(\hat{x},\hat{X})=(0,0)$. Writing the Hamilton equations and linearising around $(0,0)$ readily gives
\beq \label{HamEq}
\frac{\ud \hat{x}}{\ud t} = \frac{\partial H}{\partial \hat{X}} = 8 x_c \hat{X} + o(x,X) \,, \quad  \frac{\ud \hat{X}}{\ud t} = -\frac{\partial H}{\partial \hat{x}} = - \frac{Y_2}{x_c} \hat{x} + o(x,X) \,.
\eeq
Now, since the potential $x\mapsto Y(x)$ must be convex, (recall the construction of an orbit on figure \ref{fig:Henon}), it is clear that we must have
$Y_2\geq 0$. Consequently, the eigenvalues $(\ell_1,\ell_2)$ of the linearised system \eqref{HamEq} are
\beq
\ell_1 = \ui\sqrt{8Y_2} \quad \text{and} \quad \ell_2 = - \ui\sqrt{8Y_2} \,.
\eeq
These eigenvalues are conjugate, imaginary numbers, allowing us to conclude that the equilibrium $(\hat{x},\hat{X})=(0,0)$ is, indeed, elliptic, as our use of the Birkhoff normal form requires.

\subsubsection{Circularising the equilibrium neighbourhood}

Next, notice that the quadratic part of $H$ in \eqref{Hhat} describes ellipses in the $(\hat{x},\hat{X})$-plane. As we aim, eventually, towards a polar-like system of action-angle coordinates, we would like to \textit{circularise} these ellipses; that is, have the same coefficients in front of $\hat{x}^2$ and $\hat{X}^2$ in equation \eqref{Hhat}. This can be done easily by yet another change of variables. Explicitly, we set $(\hat{x},\hat{X})=(\eta z,\gamma Z)$ (a homothety for fixed $\Lambda$) and choose $(\eta,\gamma)$ such that: (i) the transformation is symplectic, and (ii) the coefficients in front of $z^2$ and $Z^2$ are equal in the new variables. A calculation reveals that condition (i) holds if $\gamma=1/\eta$, while condition (ii) holds if we set $\eta^4=8x_c^2/Y_2$. Expressing the Hamiltonian with the new $(z,Z)$-variables\footnote{\label{fn2}We would also need to change the angle $\hat{\theta}\mapsto\hat{\theta} + \tfrac{\eta'(\Lambda)}{\eta(\Lambda)}\hat{x}$, to ensure symplecticity in the 4D phase space.}, we find
\beq \label{Hz}
H_{\Lambda}(z,Z) = Y_1 + \sqrt{2Y_2}\bigl(z^2 + Z^2 + c_0 z Z^2+ c_1 z^3 + c_2 z^4\bigr) + o(z^4)\,,
\eeq
where we see that our phase-space ellipses have indeed been circularised, and where we defined new coefficients $(c_0,c_1,c_2)$ by
\beq \label{abc}
c_0 = \frac{8^{1/4}}{Y_2^{1/4}x_c^{1/2}}\,, \quad c_1= \frac{8^{1/4}}{3}\frac{x_c Y_3-3Y_2}{x_c^{1/2}Y_2^{5/4}} \quad \text{and}\quad c_2 = \frac{8^{1/2}}{12}\frac{12 Y_2 - 4 x_c Y_3 + x_c^2 Y_4}{x_c Y_2^{3/2}} \,.
\eeq
One more time, we emphasise that, in the complete, 4-dimensional phase space, these coefficients all depend on $\Lambda$, through $x_c(\Lambda)$ and $Y_n(x_c(\Lambda))$. Next, we simplify the $(z,Z)$-dependent part in the parentheses of \eqref{Hz}.

\subsubsection{Flowing towards the normal form}

For the moment, let us rewrite \eqref{Hz} in the form $H=Y_1+\sqrt{2Y_2}\tilde{H}+o(z^4)$, where
\beq \label{Hti}
\tilde{H}(z,Z) = z^2 + Z^2 +c_0 zZ^2 + c_1 z^3 + c_2 z^4 \,.
\eeq
As our final aim is to introduce polar-type coordinates, we would like $\tilde{H}$, a polynomial in $(z,Z)$, to be written solely as powers of $\rho=z^2+Z^2$, which would then correspond to the radial part of the polar-type coordinates. The best way to massage $\tilde{H}$ into this form is to make a transformation derived from the flow of another (polynomial) Hamiltonian $\Phi$. Let us take a moment to explain this method more clearly. \\

Using the flow of a secondary Hamiltonian can be viewed as a very general procedure to produce symplectic transformations $(z,Z) \mapsto (\bar{z},\bar{Z})$. Let $\Phi(z,Z)$ be some arbitrary Hamiltonian, and let $\phi_t$ be the \textit{flow} associated to $\Phi$, such that $\phi_t:(z,Z)\mapsto(\bar{z},\bar{Z})=(z(t),Z(t))$, where $(z(t),Z(t))$ is the solution to Hamilton's equations for $\Phi$. For $t=1$, the map $\phi:=\phi_1$ is appropriately called the \textit{time-one flow}, as it sends a point $(z,Z)=(z(0),Z(0))$ (corresponding to some initial condition $t=0$) to some other point $(\bar{z},\bar{Z})=(z(1),Z(1))$ (corresponding to its updated value at $t=1$). Choosing $\Phi$ in the right way allows one to determine the dynamics between $t=0$ and $t=1$, and thus select the image $(\bar{z},\bar{Z})$ of each point $(z,Z)$. By construction, this mapping $\phi$ defines a symplectic transformation on the phase space, because it derives from a Hamiltonian system. \\

Returning to our problem, an explicit computation adapted from \cite{FeKa.04} (with a slight adjustment for the cross term $z Z^2$ in \eqref{Hti} absent there) shows that if $\tilde{H}(z,Z)$ is of the form \eqref{Hti}, then the time-one flow $\phi$ of a well-chosen\footnote{Explicitly, $\Phi(z,Z)=b_1 Zz^2 + b_2 Z^3 + b_3 Z z^3 + b_4 z Z^3$, where $(b_1,b_2,b_3,b_4)$ are combinations of $(c_0,c_1,c_2)$. 
} $\Phi(z,Z)$ defines a set of coordinates $(\bar{z},\bar{Z})$ precisely such that the polynomial Hamiltonian \eqref{Hti} now reads, in the ``bar'' variables:
\beq \label{Hbar}
\bar{H}(\bar{z},\bar{Z}) = \bar{z}^2+\bar{Z}^2 + C\,(\bar{z}^2+\bar{Z}^2)^2 + O(5) \,,
\eeq
where $O(5)$ contains terms of order 5 or more in $(\bar{z},\bar{Z})$, and $C$ is expressed in terms of the constants appearing in \eqref{Hti}, by
\beq \label{C}
C=-\frac{3}{32} (5 c_1^2-4c_2^2+2c_1c_0 + c_0^2) \,.
\eeq
We can now go back to the original Hamiltonian \eqref{Hz} and express it in the new variables $(\bar{z},\bar{Z})$. To this end, we replace $\tilde{H}(z,Z)$ in \eqref{Hz} (recall that $H=Y_1+\sqrt{2Y_2}\tilde{H}+o(z^4)$) by $\bar{H}(\bar{z},\bar{Z})$ as given in \eqref{Hbar}. We eventually find
%
%
\beq \label{Hend}
H(\bar{z},\bar{Z}) = Y_1 + \sqrt{2Y_2}(\bar{z}^2+\bar{Z}^2) + C \sqrt{2Y_2}\,(\bar{z}^2+\bar{Z}^2)^2 +O(5)\,.
\eeq
where $C$ is a function of $x_c$ given by combining equations \eqref{C} and \eqref{abc}. Expression \eqref{Hend} is then directly amendable to a normal form, as we show in the next paragraph.

\subsubsection{Polar action-angle coordinates}\label{sec:polar}

The final step to extract the normal form of \eqref{Hend} is to promote $\bar{z}^2+\bar{Z}^2$ to an action variable. A classical technique \cite{Arn} is to think of $(\bar{z},\bar{Z})$ as a kind of cartesian-type coordinates and pass to (symplectic) polar coordinates, by setting
\beq
z = \sqrt{2\rho} \cos \varphi \quad \text{and} \quad Z = -\sqrt{2\rho} \sin \varphi \,,
\eeq
where this expression is necessary to enforce symplecticity. Inserting into equation \eqref{Hend} the new coordinates $(\rho,\varphi)$ gives us our final expression for the Hamiltonian
\beq \label{Hfinal}
H(\rho,\varphi) = Y_1 + \sqrt{8Y_2}\rho + C \sqrt{32Y_2} \,\rho^2 +o(\rho^2)\,.
\eeq
Now we can compute $C=C(\Lambda)$ in terms of $x_c$ and the $Y_n$'s from equations \eqref{C} and \eqref{abc}. The normal form $N_1(\rho)$ of \eqref{Hfinal} is therefore
\beq \label{N1}
N_1(\rho) = Y_1 + \sqrt{8Y_2}\,\rho +\frac{1}{2} \biggl( \frac{4 Y_3}{Y_2}  + \frac{x_c}{3Y_2^2} \bigl( 3Y_2 Y_4 - 5 Y_3^2 \bigr) \biggr) \,\rho^2 \,.
\eeq
It should be noted that no assumption about the radial potential $Y(x)$ has been made to derive this normal form. In particular, $Y(x)$ is not required to be isochrone, and, much like in \cite{FeKa.04}, this normal form is valid for any radial potential. 
We know turn to the derivation of the normal form for an isochrone potential.

\subsection{Birkhoff normal form: isochrone potential}\label{sec:Biso}

In general, the strength and simplicity of a normal form is usually balanced by the (analytic) complexity involved in its derivation (see e.g. the normal form of the restricted $N$-body problem, \cite{Fe04,ChPi.11}). However, in the case of an isochrone potential, things are much simpler thanks to the symmetry at play. We explain, in this subsection, how to construct the normal form of the Hamiltonian in that particular, isochrone case. The simplicity of the argument should then be compared to the previous subsection \ref{sec:Bgen}, where without the isochrone assumption the calculation was much more involved, and very close to that of \cite{FeKa.04}.\\

We will follow the same steps used in section \ref{sec:hamiso}. In particular, we start from the following result: if the potential is isochrone, the radial action $J$ decomposes into a sum of two terms, one $\xi$-dependent and one $\Lambda$-dependent, as was shown in \eqref{RA}. For convenience we rewrite this as
\beq \label{genJ}
J(\xi,\Lambda)=F(\xi)-G(\Lambda)\,,
\eeq
where $F,G$ are two functions\footnote{We assume that $F,G$ behave nicely: they can be differentiated several times and inverted on their domain of definition. We know this will be the case as we know their explicit form \eqref{T}, \eqref{Theta}.} such that $F'(\xi)=T(\xi)/2\pi$ and $G'(\Lambda) = \Theta(\Lambda)/2\pi$, since for any radial potential, \eqref{derJ} must hold. In section \ref{sec:hamiso} we had the explicit expressions of $F$ and $G$ (recall \eqref{action}) but these have been obtained in \cite{RP.20} assuming what we are attempting to prove, namely the isochrone theorem. As we shall see, these explicit forms are not required to make the computation. 

Now let us fix a value of $\Lambda$, and solve equation \eqref{genJ} for the energy $\xi$ in terms of the radial action $J$. Since that expression holds for any $\xi$, i.e. any numerical value of the Hamiltonian $H=\xi$, we have just obtained $H$ expressed in terms of the action $J$, at fixed $\Lambda$. This expression reads
\beq \label{HJL}
H(J,z_J) = F^{-1}(G(\Lambda)+J)\,,
\eeq
where we have re-introduced the dependence on the radial angle variable $z_J$ associated to the radial action $J$, for completeness. Let us emphasise one more time that, at this stage, we do not know the expressions of $F,G$. They can only be computed once the isochrone theorem is demonstrated. When this is done equations \eqref{genJ} and \eqref{HJL} will become \eqref{action} and \eqref{ham}, respectively.

As emphasised in the last section, for a given value of $\Lambda$, circular orbits are relative equilibria of $H$ and correspond to $J=0$. Let us then Taylor-expand \eqref{HJL} around $J=0$ and set $H(0,z_J):=\xi_c(\Lambda)$ as the energy of that circular orbit. We readily get
\beq
H(J,z_J) = \xi_c + \frac{1}{F'(\xi_c )} J + \frac{1}{2} \biggl(-\frac{F''(\xi_c)}{F'(\xi_c )^3} \biggr) J^2 +o(J^2) \,.
\eeq
We can now extract the normal form of the above Hamiltonian $H(J,z_j)$, which we denote by $N_1(J)$, such that $H(J,z_J)=N_1(J)+o(J^2)$. Using the property $F'(\xi)=T(\xi)/2\pi$ one more time, we find
\beq \label{N2}
N_2(J) = \xi_c + \frac{2\pi}{T(\xi_c )} J + \frac{1}{2} \biggl(-\frac{4\pi^2 T'(\xi_c)}{T(\xi_c )^3} \biggr) J^2 \,,
\eeq
where $T(\xi_c)$ is understood as the limit of $T(\xi)$ when $\xi\rightarrow\xi_c(\Lambda)$ for fixed $\Lambda$, since the radial period of a circular orbit can be ambiguous to define. Equation \eqref{N2} is, for a given $\Lambda$, a normal form for $H$, but we emphasise that it holds \textit{only if the potential is isochrone}, otherwise equation \eqref{genJ} (from which \eqref{N2} follows) does not hold in the first place. With this second normal form at hand, we can finally turn to the applications, in the next and last section.

\section{Three applications of the normal form} \label{sec:quatre}

In this fourth and last section, we use the two Birkhoff normal forms \eqref{N2} and \eqref{N1} of the Hamiltonian describing a particle in an isochrone potential. By exploiting the equality between their respective Birkhoff invariants, we provide: (1) a proof of the fundamental theorem of isochrony \eqref{thm}; (2) a proof of the Bertrand theorem; and (3) a proof of the generalised Kepler's third law \eqref{T}. These three items are presented in each of the three following subsections.

\subsection{Fundamental theorem of isochrony} \label{sec:fti}

The two Birkhoff normal forms $N_1$ and $N_2$ derived in the previous section define two sets of three Birkhoff invariants (according to \eqref{defNorm}), one for each normal form. They must be equal, by unicity of the normal form. From the first normal form \eqref{N1}, derived in the $\rho$ action coordinate, their expression is
\beq \label{Binv2}
\mathfrak{l}_1 = Y_1\,, \quad 
\mathfrak{b}_1 = \sqrt{8Y_2} \,, \quad \text{and} \quad
\mathfrak{B}_1 =\frac{4 Y_3}{Y_2} +  \frac{x_c}{3Y_2^2} \bigl( 3Y_2 Y_4 - 5 Y_3^2 \bigr) \,,
\eeq
where we emphasise that each of these invariants are $\Lambda$-dependent, through the derivatives of the potential $Y_n=Y^{(n)}(x_c(\Lambda))$ and (twice the square of) the radius of the circular orbit $x_c=x_c(\Lambda)$. The second normal form \eqref{N2} then provides an alternative expression 
\beq \label{Binv1}
\mathfrak{l}_2 = \xi_c(\Lambda) \,, \quad 
\mathfrak{b}_2 = \frac{2\pi}{T(\xi_c)} \,, \quad \text{and} \quad 
\mathfrak{B}_2 = -4\pi^2\frac{ T'(\xi_c)}{T(\xi_c)^3} \,,
\eeq
where, once gain, they are $\Lambda$-dependent through the energy of the circular orbit $\xi_c = \xi_c(\Lambda)$. The invariants \eqref{Binv1} are computed under the assumption that $Y(x)$ (or equivalently $\psi(r)$) is isochrone, while the invariants \eqref{Binv2} are valid for any $Y(x)$ (not necessarily isochrone). However, if we assume $Y(x)$ isochrone, then \eqref{Binv2} and \eqref{Binv1} are the Birkhoff invariants of the same system (a particle of angular momentum $\Lambda$ and energy $H=\xi$ in an isochrone potential). Therefore, from now on we assume that $Y(x)$ is isochrone, and derive the isochrone theorem \eqref{thm} by exploring the consequences of the three equalities $ (\lfk_1,\bfk_1,\Bfk_1)=(\lfk_2,\bfk_2,\Bfk_2)$ in three steps, one for each order of invariants.

\subsubsection{Zeroth order invariant}

The first equality $\lfk_1=\lfk_2$ provides a link between the energy of the circular orbit of angular momentum $\Lambda$ and the first derivative of $Y$, namely: 
\beq \label{Birk1}
Y'(x_c(\Lambda)) = \xi_c(\Lambda) \,.
\eeq
This equation is consistent with the construction of an orbit in the $x=2r^2$ variable, as we explained in figure \ref{fig:Henon}. Indeed, a circular orbit of energy $\xi_c$ corresponds the line $y=\xi_c x-\Lambda^2$ being tangent to the curve $Y(x)$. Therefore, their respective slope must be equal at the tangency point $x_c$, hence $\xi_c=Y'(x_c)$. The other consequence of that equation is how $\xi_c$ varies with respect to $\Lambda$. Indeed, we have 
\beq \label{prime}
\frac{\ud \xi_c}{\ud \Lambda} = \frac{\ud Y_1}{\ud \Lambda} = x_c'(\Lambda)Y_2 \,,
\eeq
where a prime denotes $\ud /\ud \Lambda$, and the Leibniz rule must be used to compute the derivative of $Y_1=Y'(x_c(\Lambda))$. Equation \eqref{prime} will be useful below. 

\subsubsection{First order invariant}

The second equality $\bfk_1=\bfk_2$ implies a relation between the radial period and the second derivative of $Y$ at $x_c$, namely 
\beq \label{Birk2}
Y''(x_c(\Lambda))=\frac{\pi^2}{2}\frac{1}{T(\xi_c(\Lambda))^2}\,.
\eeq
Once we know $Y(x)$, this equation allows to derive easily the generalisation of the Kepler's third law of motion, which we saw back in \eqref{T}. The other consequence of \eqref{Birk2} is an equation for $\Bfk_2$. Indeed, differentiating \eqref{Birk2} with respect to $\Lambda$ readily gives
\beq \label{primeprime}
x_c'Y'''(x_c)=-\pi^2 \frac{\ud\xi_c}{\ud \Lambda}\frac{T'(\xi_c)}{T(\xi_c)^3}  
\,.
\eeq
We see that \eqref{primeprime} is very similar to the expression of $\Bfk_2$ in \eqref{Binv1}. In fact, inserting \eqref{prime} in \eqref{primeprime} and comparing the resulting with \eqref{Binv1} readily gives the relation
\beq\label{primeprimeprime}
\Bfk_2=\frac{4Y_3}{Y_2} \,,
\eeq
where we used the fact that $x_c'(\Lambda)\neq 0$, which follows by differentiating \eqref{xc} with respect to $\Lambda$ to obtain $x_c' x_c Y_2=2\Lambda$. With equation \eqref{primeprimeprime} at hand we may finally complete the proof of the fundamental theorem of isochrony.

\subsubsection{Second-order invariant}

Lastly, we insert in the equality $\Bfk_1=\Bfk_2$ the expression \eqref{primeprimeprime} for $\Bfk_2$, and the expression \eqref{Binv2} for $\Bfk_1$, to conclude that
\beq \label{fun}
\frac{x_c}{3Y_2^2} \bigl( 3Y_2 Y_4 - 5 Y_3^2 \bigr) = 0 \,.
\eeq
Since $x_c=2r_c^2\neq 0$, the parenthesis must vanish. Recalling the notation $Y_n=Y^{(n)}(x_c(\Lambda))$, and since \eqref{fun} should hold for any $\Lambda$, we may now let $\Lambda$ vary continuously. By continuity of $\Lambda\mapsto x_c(\Lambda)$, the equation $3Y_2 Y_4 = 5 Y_3^2$ is nothing but an ODE for the function $x_c\mapsto Y(x_c)$, i.e., the function $Y$. Therefore, at least on some open interval of $\RR_+$, we must have
\beq \label{edo}
3 Y^{(2)}Y^{(4)} = 5\bigl(Y^{(3)}\bigr)^2 \,.
\eeq
It turns out that equation \eqref{edo} is \textit{the universal differential equation for parabolae}, in the sense that its solutions cover all and only functions $Y$ whose curve $y=Y(x)$ are parabolae in the $(x,y)$ plane. A short proof of this statement is included in appendix \ref{app:para}. This concludes the proof of the isochrone theorem \eqref{thm}. Before going to the next paragraph, let us mention that \eqref{edo} can be simply written as an ODE for the $\Lambda$-dependent Birkhoff invariants $(\lfk,\bfk,\Bfk)$ themselves, namely 
\beq \label{edoBirk}
\Bfk \, \frac{\ud \lfk}{\ud \Lambda} = \bfk \, \frac{\ud \bfk}{\ud \Lambda} \,.
\eeq
In fact we could have obtained \eqref{edoBirk} readily from the fact that $\Bfk_2 \lfk_2'=\bfk_2 \bfk_2'$ (here a prime denotes $\ud/\ud\Lambda$), which can be seen easily from \eqref{Binv1}. That the isochrone theorem follows from such a simple differential relation between the Birkhoff invariants constitutes a very nice and fundamental characterisation of isochrony. More insight on \eqref{edoBirk} is provided in appendix \ref{app:B}. 

\subsection{Bertrand theorem} \label{sec:Ber}

There is another fundamental result that we can derive from this formalism: the Bertrand theorem. As mentioned before, this was actually done in \cite{FeKa.04} and was the main motivation behind our exposition here. However, we would like to present it in the light of isochrony. Indeed: as we mentioned back in section \ref{sec:gauge}, the Bertrand theorem states that only the Harmonic and Kepler potentials generate closed and only closed orbits. But notice that both of these potentials are isochrone. Therefore, we expect the Bertrand theorem to be a corollary of the isochrone theorem (as was argued already in \cite{SPD}). We prove the Bertrand theorem in two steps: first we show that a Bertrand potential $Y(x)$ must be a power law (up to a linear term); and second, that it must be isochrone.\\

Let us consider the normal form \eqref{N1}, which holds for any radial potential $Y(x)$, including isochrone and  Bertrand potentials. Let us write the corresponding Hamiltonian $H(\rho,\phi,\Lambda,\vartheta)$ in the complete, 4D-phase space with the two pairs $(\rho,\phi)$, $(\Lambda,\vartheta)$ of action angle variables. We have seen that it reads
\beq
H(\rho,\Lambda)= \lfk_1(\Lambda) + \bfk_1(\Lambda) \rho + \frac{1}{
2} \Bfk_1(\Lambda) \rho^2 + o(\rho^2) \,,
\eeq
where $(\lfk_1,\bfk_1,\Bfk_1)$ are given in terms of $Y(x_c(\Lambda))$ in \eqref{N1}. Associated to the action variables $(\rho,\Lambda)$, the corresponding frequencies $(\omega_{\rho},\omega_{\Lambda})$ of this Hamiltonian thus read 
\beq \label{ombirk}
\omega_{\Lambda} := \frac{\partial H}{\partial \Lambda} = \frac{\ud\lfk_1}{\ud\Lambda} + o(1) \,, \quad \text{and} \quad \omega_{\rho} := \frac{\partial H}{\partial \rho} = \bfk_1(\Lambda) + o(1) \,.
\eeq
If $Y(x)$ satisfies the Bertrand theorem, then all the orbits are closed in real space. In phase space, a closed orbit corresponds to a pair of actions $(\rho,\Lambda)$ (recall figure \ref{fig:torus}) that defines a torus, on which the associated curve wraps around, but ultimately closes on itself. This is called a resonant orbit, i.e., an orbit for which there exists integers $(k_{\Lambda},k_{\rho})\in\ZZ$ such that $k_{\Lambda}\omega_{\Lambda}+k_{\rho}\omega_{\rho}=0$. Now, since \textit{each and every} orbit must be closed for a Bertrand potential, this means that these integers $(k_{\Lambda},k_{\rho})$ are actually independent of the pair $(\rho,\Lambda)$. In other words, there exists a $Q\in\QQ$ such that for all $(\rho,\Lambda)$,
\beq \label{Q}
\omega_{\Lambda} (\rho,\Lambda) = Q \, \omega_{\rho} (\rho,\Lambda) \,.
\eeq
We emphasise that equation \eqref{Q} should hold for any pair of actions $(\rho,\Lambda)$. In particular, \eqref{Q} should hold for a given $\Lambda$ in the limit $\rho\rightarrow 0$ (quasi-circular orbits). According to \eqref{ombirk}, this means that
\beq \label{ind}
\frac{\ud\lfk_1}{\ud\Lambda} = Q \, \bfk_1 \,. 
\eeq
It is rather remarkable that the Bertrand theorem is equivalent to such a simple condition, namely a differential equation for the Birkhoff invariants. We can solve this equation easily. First we insert the definitions \eqref{Binv2} of $\lfk_1(\Lambda)$ and $\bfk_1(\Lambda)$ in terms of $Y_1$ and $Y_2$. Then the calculation reads
\beq
\label{ind+}
\eqref{ind}
\,\,\Rightarrow\,\, x_c' Y_2 = Q \sqrt{8Y_2} 
\,\,\Rightarrow\,\, \Lambda^2 = 2x_c^2Q^2 Y_2 
\,\,\Rightarrow\,\, x_c Y_1 - Y = 2x_c^2Q^2 Y_2 \,,
\eeq
where in the first step we differentiated with the Leibniz rule (much like in \eqref{prime}), in the second step we squared and used $x_c'x_cY_2=2\Lambda$ which we obtain by differentiating \eqref{xc} with respect to $\Lambda$, and in the last step we used \eqref{xc} once more to remove $\Lambda$. 
Much like equation \eqref{fun} can be seen as an ODE for $x_c\mapsto Y(x_c)$, the rightmost equation in \eqref{ind+} is an ODE too, in which $Q\in\QQ$ is a parameter. The solution to this ODE is simply found as
\beq \label{Ber}
Y(x) = C_1 x  + C_2 x^K \,, \quad \text{with} \quad K:=\frac{1}{2Q^2} \,,
\eeq
with two integration constants $(C_1,C_2)\in\RR^2$. The linear term $C_1 x$ corresponds to the addition of a constant in the potential $\psi(r)$ (recall $Y(2r^2)=2r^2\psi(r)$). As it does not affect the dynamics, we leave it aside and set $C_1=0$.\\

On the one hand, we have shown that if $Y(x)$ is a Bertrand potential, then according to equation \eqref{Ber} it must be a power law. On the other hand, it is clear that a Bertrand potential must be isochrone: if all bounded orbits are closed, then the apsidal angle $\Theta(\xi,\Lambda)$ must be a constant, rational multiple of $2\pi$. In particular, as a constant function it is independent of the energy $\xi$ of the particle. But this characterises isochrony according to \eqref{defisoT}. The conclusion is thus that a Bertrand potential must, at once, have the form of a power law and that of a parabola. The only parabolae that verify this property are either the square root $Y\propto\sqrt{x}$ or the quadratic $Y\propto x^2$. In terms of the variable $r$, this means that either $\psi(r)\propto 1/r$ (the Kepler potential), or $\psi\propto r^2$ (the Harmonic potential). Moreover, according to \eqref{Ber}, these two cases correspond to $K=1/2$ and $K=2$, i.e. to $Q=1$ or $Q=1/2$, respectively. In light of the link between the apsidal angle $\Theta$ and the ratio of Hamiltonian frequencies \eqref{freq}, we recover the classical formulae \eqref{Thetas}.

\subsection{Generalisation of Kepler's Third Law} \label{sec:K3}

As a final application of the Birkhoff normal forms, let us consider once more the equality $\bfk_2=\bfk_1$, which was written explicitly in terms of $Y$ in \eqref{primeprime}. Re-arranging this equation provides, for any $\Lambda$,
\beq \label{T2}
T(\xi_c(\Lambda))^2 = \frac{\pi^2}{2} \frac{1}{Y''(x_c(\Lambda))} \,.
\eeq
But now, recall the initial definition of isochrone potentials \eqref{defiso}: the radial period should be independent of $\Lambda$. Although here the equation holds for the circular orbit of energy $\xi_c$, there exist other, non-circular orbits with the same energy. Geometrically, they can be constructed by transLating the line $y=\xi_c x-\Lambda^2$ upward on figure \ref{fig:Henon}. By construction, all these orbits (defined by the translation) only see their angular momentum change, not their energy (a translation preserves the slope). Consequently, their radial period (squared) is numerically equal to \eqref{T2}. Summarising, we can now write that an orbit of energy and angular momentum $(\xi,\Lambda)$ has a radial period $T(\xi)$ given by 
\beq \label{T3}
T(\xi) = \frac{\pi}{\sqrt{2}} \frac{1}{\sqrt{Y''(x_c(\xi))}} \,,
\eeq
where now $x_c(\xi)$ denotes the abscissa of the circular orbit with energy $\xi$, obtained by a downward translation (cf figure \ref{fig:Henon}). Equation \eqref{T3} is in complete agreement with formula (B2) derived in the appendix of \cite{RP.20}, where $T$ was expressed in terms of the radius of curvature $R_c$ of the parabola at the point of abscissa $x_c$. Recalling the link between curvature and the second derivative for explicit curves, equality between the two formulae follows. To obtain the general expression \eqref{T} in terms of $\xi$ and the parabola parameters $(a,b,c,d,e)$, one simply needs to compute the second derivative of a given parabola $Y(x)$, evaluate it at $x_c$ and insert the result in \eqref{T3} (this is explained in details in section IV.4.\textit{1} of \cite{RP.20}). The result \eqref{T} follows immediately.  
%
%
%
%
%

\section*{Conclusions}

In this article, we have explained how the notion of isochrony for radial potentials, as first introduced by Michel Hénon \cite{HeI.59,HeII.59,HeIII.59} and explored in depth in \cite{SPD,RP.20}, was most naturally expressed in the context of Hamiltonian mechanics. Most of our results provide a thorough and self-consistent answer to some questions that were left open in our previous work \cite{RP.20}. \\

In particular, in section \ref{sec:deux} we used the remarkable property of the radial action $J$ \eqref{genJ}, which, along with the angular momentum action $\Lambda$, provides a system of angle-action coordinates particularly well-suited for isochrone potentials, due to its energy and angular momentum splitting \eqref{RA}. Using these variables, we have: (1) solved at once all the dynamics of test particles in any isochrone potential in terms of a generalised eccentric anomaly \eqref{smaxis},\eqref{thetaE}; and (2) shown how the Kepler equation \eqref{Keplereq} and Kepler's third law \eqref{T} -- which are classically (and rightfully) associated to the two-body problem only -- are actually universal properties of all isochrone orbits. \\

Along with these generalisations and explicit solutions, the Hamiltonian point of view used in this paper allowed us to provide, in sections \ref{sec:trois} and \ref{sec:quatre}, a natural and self-consistent proof of the fundamental theorem of isochrony \eqref{thm} that relates isochrone potentials to parabolae in the plane. With the help of the Birkhoff normal form written around circular orbits, we have provided a proof that does not rely on abstract and unrelated mathematical results (as is the case both in \cite{SPD} and \cite{RP.20}), but only uses fundamentals of Hamiltonian mechanics -- in our case, the unicity of the Birkhoff invariants \eqref{edoBirk}. Additionally, we showed how this normal form formalism gives as an elementary by-product, the Bertrand theorem \eqref{ind} of classical mechanics, and the Kepler third law \eqref{T} generalised to all isochrone (initially derived in \cite{RP.20}).\\

From a more global point of view, the results derived in this paper show in a definitive manner that the century old, intricate symmetries associated with the harmonic and Kepler potentials [e.g., Kepler's laws of motion and Kepler's equation (cf chapter 2 of \cite{BoPuV1}, the Bertrand theorem (cf \cite{SPD} and references therein), the Bohlin-Levi-Civita transform (cf \cite{Bo.11,LyJi.08}), etc] are actually sub cases of a much larger, isochrone paradigm, as was already emphasised in \cite{SPD,RP.20}. However, whether or not these results can be applied and useful to actual physical problems is still unclear (except, of course, for the Kepler and harmonic potential). It is worth mentioning, though, that the Hénon potential has been known to model particularly well some clusters of stars (see \cite{SPD} and references therein), and seem to be related to cluster formation in the first place \cite{SPal.19}. What's more, toy-models for dark matter halos \cite{Me.06} using the Hollowed class as well as quark confinement \cite{Mu.93} using the Bounded class could also prove to be useful and should be explored. Lastly, stability analysis of perturbed isochrone potentials may be a way to make progress towards the problem of dark matter halo collapse and virialisation \cite{SPal.19}. \\

Another thing that is missing in our analysis, but would be interesting to investigate, is an explicit geometrical construction of this generalised eccentric anomaly. In the Kepler case, this is well-known (see, e.g., section 2.1.2 of \cite{Arn}), and much like everything else regarding isochrony, there may be a ``straightedge and compass'' construction for the eccentric anomaly of a general isochrone orbit. Another line of research worth mentioning is a formula that encompasses both the non-harmonic and harmonic case (compare \eqref{smaxis} and \eqref{thetaE} with \eqref{Hasol} and \eqref{thetaharm}, respectively). We leave the resolution of these two geometric problems for future work. \\

As a final word, we would like to stress the remarkable fact that all and any isochrone question can be answered with explicit and analytical equations, and that this is of particular interest for pedagogical and academic purposes. As we already mentioned in \cite{RP.20}, we think that mathematical physics problems (such as isochrony) may be hard to find in the evermore specialising literature. This being said, we encourage the interested reader to use all these isochrone results to illustrate the power of Poincaré's ``geometrical thinking'': to simplify and solve a differential problem that is complex at first sight, using nothing but symmetries and geometry, be it Euclidean as in \cite{RP.20}, or symplectic, as we proposed in the present paper.

\acknowledgments

PR is grateful and indebted to J.~F{\'e}joz for many helpful discussions on the theory of Birkhoff normal forms. 

\clearpage


%
\begin{table}[!ht]
    \caption{List of frequently used symbols in \cite{SPD,RP.20} and this paper.}
    \vspace{0.2cm}
	\begin{tabular}{ccc}
		\toprule
		\textbf{Symbol}                         & \textbf{Description}          & \textbf{Definition}  \\
		\midrule
		\textbf{Classical mechanics}            &                                           & \\
		$\xi$                                   & (mechanical) energy,                      & \\ 
		$\Lambda$                               & angular momentum                         & \\ 
		$T$                                     & radial period                             & \eqref{defT}\\ 
		$\Theta$                                & apsidal angle                             & \eqref{defTheta}\\ 
		$\psi(r)$                               & radial potential                          & \\ 
		$(r,\theta)$                            & polar coordinates                         & \\
		\midrule
		\textbf{Isochrony potentials}           &                                           & \\
		$x=2r^2$                                & Hénon variable                            & \\ 
		$Y(x)=x\psi(r(x))$                      & radial potential in Hénon's variable      & \\ 
		$(a,b,c,d,e)$                           & Latin parameters                          & ~\eqref{parabolaimplicit}\\ 
		$\delta=ad-bc$                          & parabola discriminant                     & \\ 
	    $x_v$                                   & abscissa of vertical tangent              & \eqref{otr}         \\ 
	    $\mu,\beta$                             & mass and length parameter                  & \cite{RP.20}, (3.12)\\
		\midrule
		\textbf{Isochrone orbits}               &                                           & \\ 
		$E$                                     & eccentric anomaly                         & ~\eqref{KE}\\ 
		$\Omega=2\pi/T$                         & radial frequency                          & ~\eqref{Omegepsi}\\ 
	    $\epsilon$                              & eccentricity                              & ~\eqref{Omegepsi}\\ 
	    $\alpha$                                & semi-major axis                           & ~\eqref{smaxis}\\ 
	    \midrule
		\textbf{Hamiltonian mechanics}          &                                           & \\ 
		$H$                                     & Hamiltonian\footnote{or perhaps should one say, the ``huygensian''. Indeed, from his own writing in the second edition of his masterwork ``Mécanique Analytique'' (1811), Lagrange introduced the letter $H$ for the ``vis viva constant'', i.e., what we nowadays call the total mechanical energy. At that time, Hamilton was only 5 years old, and it is probable that Lagrange chose $H$ for Huygens, making clear mention of him throughout his work, in particular of the insight he must have had to understand that the ``vis viva'' was conserved. More on this fascinating story can be found in \cite{image_des_maths} or appendix B of \cite{piz}, and references therein.} & \\
		$J$                                     & radial action                             & \eqref{j}\\
		$N_1,N_2$                                     & normal forms of $H$                             & \eqref{defNorm}\\
		$(J,\Lambda,z_J,z_\Lambda)$             & action-angle coordinates                 & \\
		$(\rho,\Lambda,\varphi,\vartheta)$      & action-angle coordinates                  & \eqref{Hfinal}\\
		$\lfk,\bfk,\Bfk$               & Birkhoff invariants                       & \eqref{defNorm}\\
        \bottomrule
        \vspace{5mm}
	\end{tabular}
    \label{Table}
\end{table}

\appendix

\section{Keplerian and Harmonic dynamics}\label{app:KeHo}

\subsection{Keplerian dynamics}

The dynamics in the Kepler potential $\psi(r)=-\mu/r$ is most easily found by writing an ODE for the function $u(\theta)$ where $u=1/r$ is the Binet variable (see \cite{Arn}). This ODE is linear in the case of the Kepler potential and reads $u''+u=\mu/\Lambda^2$, with $u'=\ud u/\ud\theta$. The solution for $r(\theta)$ is then easily obtained, and using $(r(0),\theta(0))=(r_p,0)$ as initial conditions gives  
\beq \label{elemts}
r(\theta)= \frac{p}{1+ \epsilon \cos\theta}\,, \quad \text{where} \quad p = \frac{\Lambda^2}{\mu} \,,\quad \epsilon = \sqrt{1+\frac{2\xi \Lambda^2}{\mu^2}} \,,
\eeq
where $p>0$ is the semi-latus rectum and $\epsilon\in[0;1[$ the eccentricity. In celestial mechanics, the angle $\theta$ is called the true anomaly. One can also parametrize the orbit with the so-called eccentric anomaly $E$, such that 
\beq\label{elements}
r(E) = \alpha(1-\epsilon \cos E) \,, \quad \text{where} \quad \tan\frac{E}{2} = \sqrt{\frac{1+\epsilon}{1-\epsilon}} \tan \frac{\theta}{2} \, ,
\eeq
where $\alpha=p/(1-\epsilon^2)=-\mu/2\xi$ is the semi-major axis of the elliptic orbit. Contrary to the true anomaly, the eccentric anomaly $E$ can be linked analytically to the orbital time $t$ elapsed along the orbit, as encoded in the Kepler equation
\beq \label{Keplereq}
\frac{\sqrt{\mu}}{\alpha^{3/2}} t = E-\epsilon \sin E 
\,,
\eeq
where it is assumed that $E=\theta=0$ at $t=0$, as our initial conditions require. Finally, Kepler's third law of motion comes as a corollary of \eqref{Keplereq}. By definition, every Keplerian orbit is an ellipse, thus the radial period $T$ coincides with the orbital period, and corresponds to $E=2\pi$. Consequently, \eqref{Keplereq} implies equation \eqref{Keplereq} since the semi-major axis $\alpha$ is linked to the energy $\xi$ by $\alpha=-\mu/2\xi$, as follows from \eqref{elemts}.

\subsection{Harmonic dynamics}

In an harmonic potential, the orbit is an ellipse, except that the origin of coordinates is at the centre of the ellipse, not at one of its focii (like the Keplerian case). The easiest way to solve the equations of motion is through cartesian coordinates. This is well-known and derived in most classical textbooks (e.g., section 3.1.(a) of \cite{BiTr}). It was mentioned at the end of section \ref{sec:one} that an isochrone potential $\psi$ always belongs to one of the four families $(\psiHa,\psiHe,\psiBo,\psiHo)$ up to a gauge-term of the form $\varepsilon + \tfrac{\lambda}{2r^2}$, where $(\varepsilon,\lambda)\in\RR^2$. The analytical results derived in this paper are general enough to take into account this gauge-liberty, for all non-harmonic potentials. For the (non-gauged) harmonic potential, the results of the last section is very classical and applies. There only remains the case of gauged-harmonic potentials, of the form \eqref{har}. Let us solve the dynamics in this general potential. 

\subsubsection{Hamiltonian in action-angle variables}

The computation of the radial action $J$ for the harmonic class is easily found by setting $b=0$ in the expressions \eqref{T} for $T$ and \eqref{Theta} for $\Theta$. They are given by 
\beq \label{TThetahar}
T = \frac{\pi}{2}\frac{\sqrt{-d}}{a}  \quad \text{and} \quad \Theta = \frac{\pi\Lambda}{\sqrt{\Lambda^2-e/d}} \,,
\eeq
where we recall that $d<0$ for an orbit to exist (otherwise the parabola is not convex) and $a>0$ since by assumption $\delta=ad>0$. From the above formulae, the expression \eqref{RA} of $J$ as a sum of integrals over $T$ and $\Theta$ can be easily turned into
\beq \label{Jhar}
J(\xi,\Lambda) = J_o + \frac{\sqrt{-d}}{4a}\,\xi - \frac{1}{2} \sqrt{\Lambda^2-\frac{e}{d}} \,,
\eeq
where $J_o$ is an integration constant that depends on $(a,b,c,d,e)$. Since we must have $J=0$ in the case of a circular orbit, a calculation (see e.g. section IV.A.\textit{1} of \cite{RP.20}) gives $J_o=-\tfrac{c}{4a\sqrt{-d}}$. Now, as in section \ref{sec:hamiso}, we get the expression of the Hamiltonian $H(J,\Lambda)$ by simply solved the above formula for the energy $\xi$, giving 
\beq \label{Hhar}
H(J,\Lambda) = -\frac{c}{d} + \frac{4a}{\sqrt{-d}} \,J + \frac{2a}{\sqrt{-d}} \sqrt{\Lambda^2-\frac{e}{d}} \,.
\eeq
The Hamiltonian of the isochrone class is therefore always linear in $J$, and linear in $\Lambda$ only in the case $e=0$, i.e., when the parabola crosses the origin of the $(x,y)$-plane (cf equation \eqref{parabolaimplicit}). Naturally, we recover the gauge term $-c/d$ (energy shift) of the harmonic potential \eqref{har}. The Hamiltonian frequencies $(\omega_J,\omega_{\Lambda})$ read in the present, harmonic case
\beq
\omega_J := \frac{\partial H}{\partial J} = \frac{4a}{\sqrt{-d}} \quad  \text{and} \quad  \omega_\Lambda := \frac{\partial H}{\partial \Lambda} = \frac{2a}{\sqrt{-d}} \frac{\Lambda}{\sqrt{\Lambda^2-e/d}} \,.
\eeq
Comparing these frequencies with \eqref{TThetahar} shows that the ratio $\omega_{\Lambda}/\omega_J$ coincides with $\Theta(\Lambda)/2\pi$, as discussed in section \ref{sec:hamiso} in the non-harmonic isochrone case. 

\subsubsection{Explicit polar solution}

First, let us re-write the equation of motion \eqref{eomx} with the explicit form \eqref{har}. We have 
\beq \label{jHa}
\frac{1}{16} \biggl( \frac{\ud x}{\ud t} \biggr)^2 = \bigr(\xi +\frac{c}{d}\bigl)x + \frac{e}{d} - \Lambda^2 + \frac{a^2}{d}x^2 \,,
\eeq
and we suppose as always that $x(t=0)=x_p$. The right-hand side of \eqref{jHa} vanishes at $t=0$ when $x=x_p$ (periastron) and at $t=T/2$ when $x=x_a$ (apoastron). Both $x_p,x_a$ are easily found as the quadratic roots of the right-hand side. Between the two, i.e., for $t\in]0;T/2[$, it is strictly positive. Therefore, for we can factorise the right-hand side of \eqref{jHa} and rearrange the result into\footnote{For the harmonic case $b=0$, we necessarily have $d<0$, since $y=Y(x)$ as given by \eqref{har} must be convex for the physical orbit to be well-defined.}
\beq \label{jHa2}
\frac{\ud x}{\sqrt{(x-x_p)(x_a-x)}} = 4\sqrt{\frac{a^2}{-d}} \ud t  \,.
\eeq
Now we integrate this equation using the Euler substitution $x\mapsto\chi$ defined by $\chi^2=\tfrac{x-x_p}{x_a-x}$. Integrating in the $\chi$ variable and going back to $x$ then gives the simple formula
\beq \label{Hasol}
x(t) = x_a + (x_p-x_a) \cos(\Omega t)^2\,, \quad \text{where} \quad \Omega=\sqrt{\frac{4a^2}{-d}} = \frac{2\pi}{T}\,,
\eeq
in which we fixed the integration constant by requiring $x(0)=x_p$. As a verification, we see that $x(t=T/2)=x_a$, in agreement with Kepler's generalised third law \eqref{T} for $b=0$ (harmonic class). Equation \eqref{Hasol} gives the solution for the radial part of the dynamics $r(t)$ via $x=2r^2$. For the angular part, we may use the angular equation of motion $\dot{\theta}=\Lambda/r^2$ and integrate with respect to $t$. We readily obtain:
\beq \label{int2}
\theta(t) = 2\Lambda \int_0^t \frac{\ud \tau}{x_a + (x_p-x_a) \cos(\Omega \tau)^2} \,.
\eeq
Let us now set $\phi=\Omega \tau$ and $\epsilon^2=1-x_p/x_a$ (such that $0<\epsilon<1$) in \eqref{int2}, and integrate with respect to $\phi$ using a partial fraction decomposition (same technique as around equation \eqref{int}). We then find the explicit expression:
\beq \label{thetaharm}
\theta(t) = \frac{2\Lambda}{\Omega x_p x_a} \sum_{\pm} \arctan \biggl( \sqrt{\frac{1+\epsilon_{\pm}}{1-\epsilon_{\pm}}} \tan \frac{\Omega t}{2} \biggr) \,, \quad E=\Omega t \,. 
\eeq
where we used the identity $x_a^2(1-\epsilon^2)=x_p x_a$ and set $\epsilon_{\pm}=\mp \epsilon$. Equations \eqref{Hasol} and \eqref{thetaharm} are true of any isochrone potential of the harmonic class. They should be  compared to their non-harmonic equivalent \eqref{xE} and \eqref{thetaE}, respectively. Based on the fact that the Kepler equation \eqref{Keplereq} reduces to $\Omega t=E$ in the limit $b\rightarrow 0$, there probably exists a way to gather all these isochrone results (both harmonic and non-harmonic) under the same formulation. We have not managed to find such formulae, but encourage the interested reader to give it a try. It would provide a strong proof of universality to the isochrone paradigm. 
As a perspective, we mention the following formulae for arctangent which is seldom found in the literature and which may be of some help. Consider the classical trigonometric identity: $\tan(x+y)=\frac{\tan x+\tan y}{1-\tan x \tan y}+\kappa\pi$ with $\kappa\in\{-1,0,1\}$ depending on where $\tan(x+y)$ lies in $[-\pi;\pi]$. Setting $(X,Y)=(\tan x,\tan y)$ we readily get the other well-known identity (assuming $\kappa=0$)
\beq \label{UV}
\arctan \frac{X+Y}{1-XY}  = \arctan X + \arctan Y \,.
\eeq
Now set $2U:=X+Y$ and $2V:=1-XY$, so that we can express $X$ and $Y$ in terms of $(U,V)$ by solving a quadratic equation. Noticing that the left-hand side of \eqref{UV} is $\arctan U/V$, we obtain a formula for the arctangent of a quotient
\beq \label{id}
\arctan \frac{U}{V} = \sum_{\pm}\arctan(U\pm\sqrt{U^2+2V-1})
\eeq
Noticing that each formula for $\theta(E)$ (\eqref{thetaE} non-harmonic and \eqref{thetaharm} for harmonic) always involves terms of the form \eqref{id}, it may be a possible starting point to find the common point between the harmonic and non-harmonic results. We leave this for future work.

\section{Analytic continuations} \label{app:complex}

\begin{figure}[!htbp]
\centering
	\includegraphics[width=.85\linewidth]{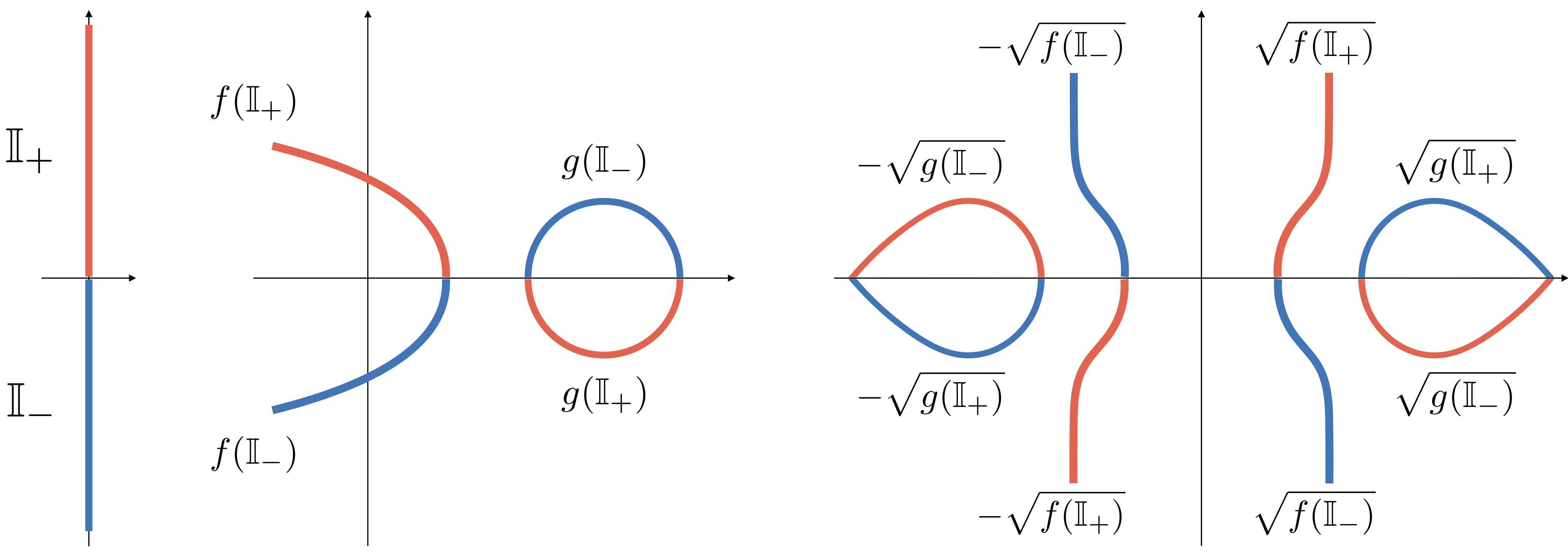}
	\caption{Successive images of the imaginary axis $\II=\II_{+}\cup\II_{-}$ under $f,g$ and then $\sqrt{f},\sqrt{g}$. Images for both branches $\pm\sqrt{\phantom{i}}$ of the (complex) square root are represented. \label{fig:mob}}
\end{figure}

The explicit formula \eqref{thetaE} for $\theta(E)$ has been obtained for all (non-harmonic) isochrone potentials with $x_v<0$. In particular, the derivation does not hold a priori for the Bounded and Hollowed class, for which $x_v>0$. Indeed, if $x_v>0$, then $\zeta^2=-x_v/2\alpha^2$ is negative, and thus $\zeta$ is imaginary (see between \eqref{int0} and \eqref{int}). However, note that the final formula \eqref{thetaE} is a sum of two terms, one with $+\zeta$ and another with $-\zeta$ (recall that $\epsilon_{\pm}=\epsilon/(1\pm\zeta)$ there). Since $\zeta\in\II\subset\CC$, this means that equation \eqref{thetaE} reads $\theta\propto F(\zeta) + F(\bar{\zeta})$ where the function $F$ is simply 
\beq \label{F}
F:z\in\CC\mapsto \frac{\epsilon}{\sqrt{(1 + z)^2-\epsilon^2}} \arctan \biggl( \sqrt{\frac{1+ z+\epsilon}{1+ z-\epsilon}} \tan \frac{E}{2} \biggr) \,,
\eeq
with $(\epsilon,E)\in [0;1] \times [0:\pi]$ seen as fixed parameters here. Now if $F$ is holomorphic around $II$ (and since its restriction to real $z$ is real-valued) then we automatically have $F(z)+F(\bar{z})=F(z)+\bar{F(z)}$ from standard results of complex analysis. But up to multiplicative positive constants $F$ can be written as $F(z)=\sqrt{f(z)} \arctan(\sqrt{g(z)})$, where 
\beq
f:z\mapsto (1 + z)^2-\epsilon^2 \quad \text{and} \quad g:z\mapsto \frac{1+ z+\epsilon}{1+ z-\epsilon} \,,
\eeq
Now let $\II=\ui\RR$ be the imaginary axis and $\II_{\pm}=\ui\RR_{\pm}$ be the lower ($-$) and upper ($+$) part of the imaginary axis. By a direct calculation, the image of $\II_+$ (resp. $\II_-$) under $f$ is the upper (resp. lower) part of a parabola, and under $g$, it is the upper (resp. lower) part of a circle (Möbius transformation). Under $z\mapsto\sqrt{z}$, and irrespective of the chosen principal value, the parabola $f(\II)$ is then mapped to a set of two disconnected curves that are complex conjugate to one-another (corresponding to the two $f(\II_+)$ and $f(\II_-)$ parts). The same is true for the circle $g(\II)$. In particular, $\sqrt{g(\II)}$ is a closed curve on the right-half plane (for the $+\sqrt{\phantom{i}}$ branch) that does not intersect the imaginary axis, such that its (complex) arctangent is well-defined and holomorphic (square roots and inverse trig functions can all be defined in terms of the complex logarithm, with which it is easy to check that all is well-defined and holomoprhic). A summary of all this is depicted on figure \ref{fig:mob}. The conclusion is that the function $F$ defined in \eqref{F} is holomorphic on $\II$, and therefore $F(z)+F(\bar{z})=2\text{Re}(F(z))$, making the formula for $\theta(E)$ also real-valued and well-defined, even in the case $x_v>0$, i.e., for Bounded and Hollowed potentials. 

\section{Universal ODE for parabolae} \label{app:para}

In this appendix, we solve the so-called universal ODE for parabolae $3 Y'' Y'''' = 5 (Y''')^2$. This ODE was already used in appendix B of \cite{SPD} to characterise parabolae. We start by the case where $Y'''=0$ which clearly is a solution. Then $Y''$ is a constant function and, therefore, $Y(x)$ is a quadratic polynomial. This corresponds to the harmonic class of parabolae \eqref{har}. If $Y'''\neq 0$, then re-arranging the equation yields $Y''''/Y'''=\tfrac{5}{3}Y'''/Y''$, which can be readily integrated as $ Y'''(Y'')^{-5/3}=C$ where $C\in\RR$. From this which we directly get $-\tfrac{3}{2}(Y'')^{-2/3}=Cx+D$ with $D\in \RR$. This implies that $Y''(x)\propto 1/(Cx+D)^{3/2}$, and therefore $Y$ is of the form \eqref{otr}, encompassing all non-harmonic types of parabolae. Reciprocally, each parabola is a solution of the ODE, which finishes the proof.

\section{More on Birkhoff invariants and action-angle transformations} \label{app:B}

In section \ref{sec:trois} and \ref{sec:quatre}, we have used the Birkhoff normal form for a 1-dimensional system (2-dimensional phase space), and thus worked with scalar Birkhoff invariants. In this section, we would like to provide an alternative way of looking at the relation between these invariants. In particular, we consider the 2-dimensional point of view of the problem (4-dimensional phase space), and consider other, more general, types of invariants. In particular, this will allow us to understand more deeply the unicity of the Birkhoff invariants. \\

Let $\vec{\omega}$ be the frequency vector made of the two natural frequencies associated with some Hamiltonian $\mathcal{H}(A, B)$, given in terms of some action variables $(A, B)$ (we are not interested in their respective angles here). It is given by
\beq \label{freqvec}
\vec{\omega} = 
\biggl(\frac{\partial \mathcal{H}}{\partial A}, \frac{\partial \mathcal{H}}{\partial B}\biggr) \,.
\eeq
Let us know construct simple quantities using $\vec{\omega}$ whose value remain unchanged under a transformation from one set of action-angle variables to another. We will use the fact that transformation between sets of action-angles is not arbitrary. Indeed, when going from a set $(A,B)$ to another, say $(A',B')$, symplecticity imposes that the old and new actions must be related by a matrix $M\in\text{SL}(2,\ZZ)$. These are $2\times 2$ matrices with determinant $1$ and coefficients in $\ZZ$. Roughly speaking, this is because angles must be transformed so that they remain angles, i.e. make $\ZZ$-linear combinations of them and not mix them with actions. Then symplecticity imposes that the actions be transformed similarly. For more on these action-angle transformations, we refer to the very clear exposition \cite{An.81}, and to the book \cite{Ma.16} for more technical details (see around proposition (6.5.3) there). Summarising, we must have
\beq \label{chg}
\begin{pmatrix} A' \\ B'\end{pmatrix} =
\begin{pmatrix} m & p \\ n & q \end{pmatrix} 
\begin{pmatrix} A \\ B \end{pmatrix}
\,, \quad \text{where} \quad \left\{
    \begin{array}{ll}
        (m,n,p,q)\in \ZZ^4 \,,\\
        mq-np=1 \,.
    \end{array}
\right.
\eeq
The change of actions \eqref{chg} induces a change in the Hamiltonian such that the frequency vector \eqref{freqvec} is transformed as $\vec{\omega}\mapsto M^ {\top}\vec{\omega}$, where $M^{\top}$ is the transpose of $M=(\begin{smallmatrix} m & p \\ n & q \end{smallmatrix})$. From then, it is easy to construct invariants by taking advantage of the fact that $\det M =1$. For example, consider the following scalar quantities
\beq \label{inv}
\mathcal{J}:= \frac{\partial \vec{\omega}}{\partial A} \wedge \vec{\omega} \,, \quad \mathcal{G}:= \vec{\omega} \wedge \frac{\partial \vec{\omega}}{\partial B} \quad \text{and} \quad \mathcal{T}:= \frac{\partial \vec{\omega}}{\partial A} \wedge \frac{\partial \vec{\omega}}{\partial B} \,,
\eeq
where $\wedge$ denotes the usual determinant between two vectors. Then the transformation \eqref{chg} leaves $\mathcal{J},\mathcal{G},\mathcal{T}$ invariant. Indeed, consider $\mathcal{J}'$, the expression of $\mathcal{J}$ in the new action variables. Then 
\beq
\mathcal{J}' = \frac{\partial (M\vec{\omega})}{\partial A} \wedge (M\vec{\omega}) = (\det M)^2 \frac{\partial \vec{\omega}}{\partial A} \wedge \vec{\omega} = \mathcal{J} \,,
\eeq
where we used in the first equality $\vec{\omega}'=M \vec{\omega}$, in the second the fact that $M$ has (constant) coefficients in $\ZZ$ and in the third $\det M=1$. A similar computation holds for both $\mathcal{G}$ and $\mathcal{T}$. Since $\vec{\omega}\wedge\vec{\omega}=0$, the three quantities \eqref{inv} are the most simple scalars built out of $\vec{\omega}$ that are invariant under \eqref{chg}. The link between $(\mathcal{J},\mathcal{G},\mathcal{T})$ (functions of $(A,B)$) and the Birkhoff invariants used in section \ref{sec:trois} and \ref{sec:quatre} is easily obtained as follows. Setting $(A,B)=(I,\Lambda)$ where $\Lambda$ is the angular momentum action and $I=\rho$ or $J$. Then if the Hamiltonian is in a normal form of the type $N(I,\Lambda)= \lfk(\Lambda)+ \bfk(\Lambda)I +\frac{1}{2} \Bfk(\Lambda) I^2 $, the quantities $(\mathcal{J},\mathcal{G},\mathcal{T})$ are easily found to be 
\beq
\mathcal{J}= \lfk\,' \Bfk - \bfk \bfk'\,,\quad \mathcal{G}= \lfk\,' \bfk' - \bfk \lfk\,'' \quad \text{and} \quad \mathcal{T}= \lfk\,'' \Bfk - \bfk' \bfk'\,.
\eeq
The quantity $\mathcal{T}(A,B)$ in \eqref{inv} is the torsion of the torus $(A,B)$. The vanishing of $\mathcal{T}$ and $\mathcal{G}$ is a necessary and sufficient condition for the Bertrand theorem to hold, as explained in \cite{FeKa.04}. We see from equation \eqref{edoBirk} that, in fact, the vanishing of $\mathcal{J}$ is a necessary and sufficient condition for the isochrone theorem to hold. What's more, it is clear from their definition \eqref{inv} that if both $\mathcal{T}$ and $\mathcal{G}$ vanish, then so does $\mathcal{J}$, since the $\mathcal{T}=0=\mathcal{G}$ implies that $\partial_B\vec{\omega}$ is parallel to both $\partial_A \vec{\omega}$ and $\vec{\omega}$, consequently $\partial_A \vec{\omega}$ is parallel to $\vec{\omega}$ and thus $\mathcal{J}=0$. Physically, this parallelism relations between the frequency vectors encodes the remarkable fact that Bertrand potentials are necessarily isochrone. 

\bibliography{main.bib}

\end{document}